\pdfoutput=1
\documentclass[%
reprint,
 amsmath,amssymb,
 aps,
]{revtex4-1}

\usepackage{graphicx}
\usepackage{dcolumn}
\usepackage{bm}
 \usepackage{subfigure}
 \usepackage{overpic}
\usepackage[mathlines]{lineno}
\begin{document}

\title{Higgs search at ATLAS}%

\author{Sofia Maria Consonni}
 \affiliation{Universit\`{a} degli Studi di Milano}
 \affiliation{INFN, Sezione di Milano}
 \email{sofia.consonni@mi.infn.it}
\collaboration{on behalf of the ATLAS Collaboration}

\date{\today}

\begin{abstract}
In this paper the results of Higgs searches at the ATLAS experiment, operating at the LHC, are summarised. The results are based on data samples corresponding to an integrated luminosity of up to 20.7 $\mathrm{fb}^{-1}$ at a centre-of-mass energy of 8 TeV and $4.6 \, \mathrm{fb}^{-1}$ at 7 TeV. The observation of a Higgs boson is established. The mass of the new boson is measured to be $125.5 \pm 0.2 \, \mathrm{(stat)}^{+0.5}_{-0.6}\,  \mathrm{(syst) \,GeV}$, and the ratio of the observed number of events and expected from a Standard Model Higgs Boson is compatible with unity, $\mu = 1.30 \pm 0.13 \, \mathrm{(stat)} \pm 0.14 \, \mathrm{(sys)}$. The spin and parity properties as well as the couplings structure are compatible with those predicted for a Standard Model Higgs Boson. No evidence of further Higgs states is found.
\end{abstract}

\maketitle

\section{\label{sec:level1} Introduction}
In the Standard Model of particle physics (SM) the Brout-Englert-Higgs mechanism allows to give masses to all massive elementary particles \cite{PhysRevLett.13.321, PhysRevLett.13.508, PhysRevLett.13.585}. The mechanism predicts as well the existence of a new scalar massive state, the Higgs boson. 
On July  $4^{\mathrm{th}}$ 2012 the ATLAS and CMS collaborations announced the discovery of a new boson in the context of the Higgs Boson search analyses \cite{Aad20121, Chatrchyan:1471016}. Since then the focus is on the study of the properties of this new particle. Moreover further decay channels and beyond the Standard model states has been continued.
This paper summarises the results obtained by the ATLAS experiment, operating at the LHC, up to Winter 2013. The results are based on data samples corresponding to an integrated luminosity up to $ 20.7 \, \mathrm{fb}^{-1}$ at a centre-of-mass energy of $8 \, \mathrm{TeV}$ and $4.6 \, \mathrm{fb}^{-1}$ at $7 \, \mathrm{TeV}$. 
The paper is organized as follows: in Section \ref{sec:higgs_pheno} a brief introduction on how the SM Higgs is expected to look like is given. In Section \ref{sec:atlas} the ATLAS detector is described, in Section \ref{sec:sm_analyses} the results of the measurements of the newly discovered boson, as well as the searches for other decay channels are summarised. Finally Section \ref{sec:bsm_searches} outlines three important Beyond the Standard Model (BSM) searches.

\section{\label{sec:higgs_pheno} Higgs boson phenomenology introduction}
The Higgs sector in the SM is fully predicted once the mass of the Higgs boson is fixed. In particular the production cross-sections and the decay branching ratios can be calculated. 

\begin{figure}[h]
\includegraphics[scale=.3]{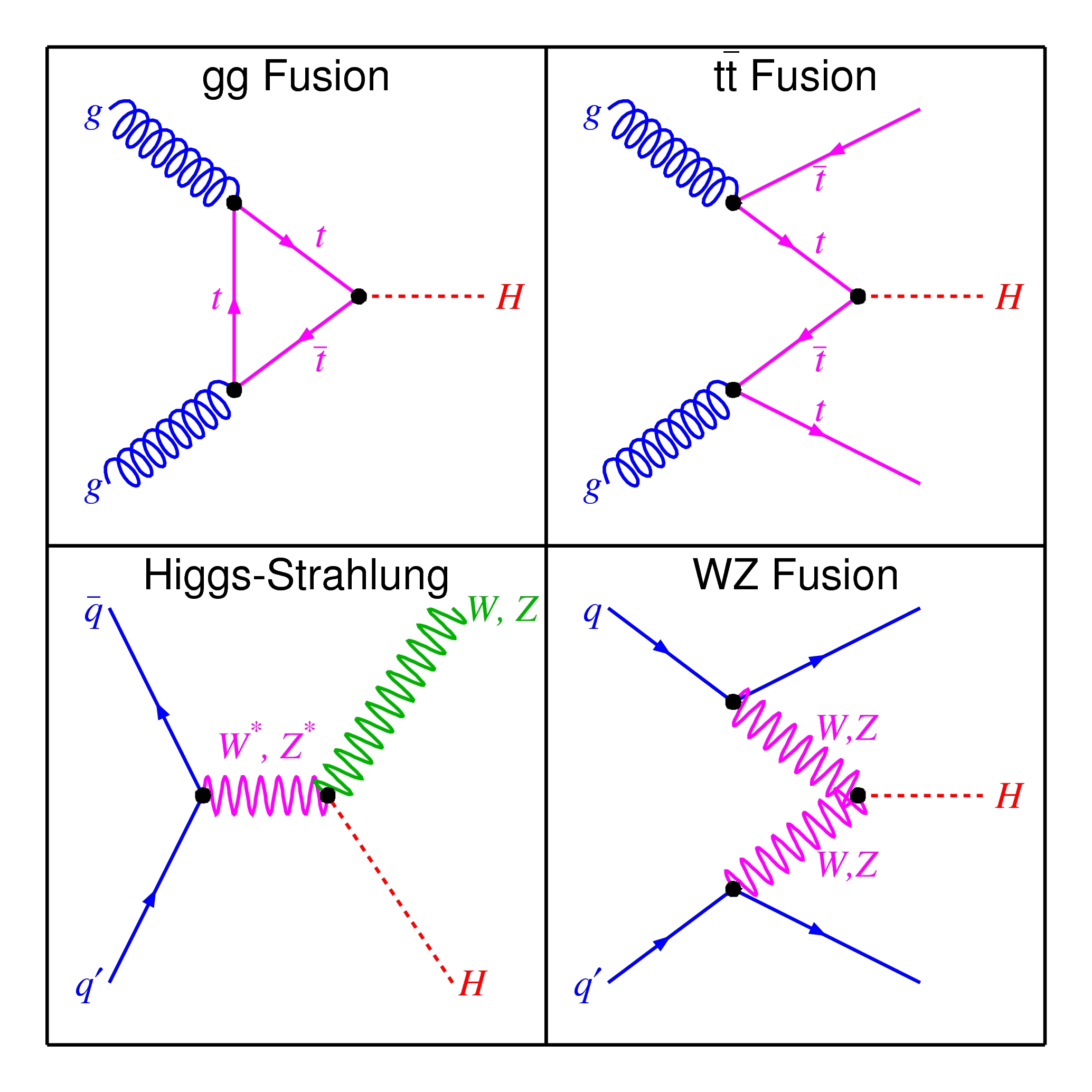}
\caption{\label{fig:prod_process} Sketch of the main Higgs production processes at the LHC.}
\end{figure}

At the LHC there are four main production mechanisms, that are sketched in Figure \ref{fig:prod_process}. The one with largest cross-section is the gluon-gluon fusion process (ggF), which proceeds through quantum loops, followed by the vector boson fusion process (VBF), which has a very distinctive signature given by two forward jets. Then immediately follows the vector boson associated production (VH), where the Higgs is produced in association with a W or a Z. Finally the $t\overline{t}$ associated production ($\mathrm{t\overline{t}H}$) is a very rare process, but with a very complex and characteristic signature.
The different production channels are used in analyses both to enhance the sensitivity, for example by targeting specific, very distinctive signature processes, or to obtain further information about couplings.

\begin{figure}[h]
\includegraphics[scale=.3]{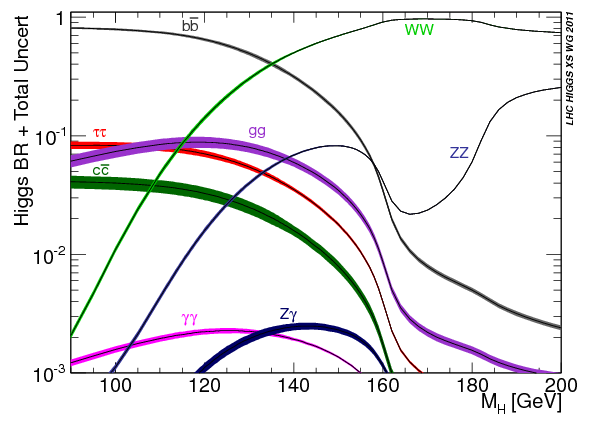}
\caption{\label{fig:branching_ratios} Higgs decay branching ratios as a function of Higgs mass \cite{Dittmaier:2011ti}.}
\end{figure}

The Higgs decay branching ratios are plotted as a function of the Higgs mass $m_{H}$ in Figure \ref{fig:branching_ratios}. At $m_{H} = 125 \,  \mathrm{GeV}$ a lot of different decay channels are open. The bosonic channels are not those with the highest branching ratios, but they benefit experimentally from a clear signature given by the presence of final state light leptons (e/$\mu$) or photons ($\gamma$). The fermionic channels, which are favoured in terms of branching ratios, are characterised by hadronic signatures (b-jets or hadronic tau decays) which are experimentally challenging. 

\section{\label{sec:atlas} The ATLAS detector}
The ATLAS detector \cite{Aad:2008zzm} is a general purpose detector operating at the LHC. It comprises different layers of detectors. The inner tracking system is closest to the interaction point, and is immersed in a 2T magnetic field provided by a superconducting solenoid. The inner detector is surrounded by high-granularity liquid-argon sampling electromagnetic calorimetry. Outside the inner detector, but covering the same rapidity region, the electromagnetic calorimeter is radially segmented in three layers, the first of which has a very fine segmentation dedicated to $\pi^{0}/\ \gamma$ separation. A thin presampler layer, covering the pseudorapidity interval $|\eta| < 1.8$, is used to correct for energy losses before the calorimeter.
Hadronic coverage is provided by an iron-scintillator/tile calorimeter in the central region ($|\eta| < 1.7$) and LAr detector in the more forward regions. The muon spectrometer (MS) surrounds the calorimeters and consists of three large air-core superconducting magnets providing a toroidal field. The combination of all these systems provides charged particle measurements together with efficient and precise lepton and photon measurements in the pseudorapidity range $|\eta| < 2.5$. Jets and missing transverse momentum ($E_{T}^{\mathrm{miss}}$) are reconstructed using energy deposits over the full coverage of the calorimeters, $|\eta| < 4.9$.

\section{\label{sec:atlas} Statistical method}

The statistical analysis for all the channels considered and for the combination of channels is based on the profile likelihood method, which is described into detail in Ref.~\cite{ATLAS-CONF-2013-014} and references therein. 
For each production mode~$i$, a signal strength factor $\mu_{i}$
defined as $\mu_{i} = \sigma_{i}/\sigma_{i, SM}$ is introduced. Similarly, for each decay final state $f$, a factor $\mu_{f} = B_{f}/B_{f, SM}$ is
introduced. For each analysis category $k$ the number of signal events $n_{\mathrm{signal}}^{k}$ is parametrized as

\begin{equation}
n_{\mathrm{signal}}^{k} = \left( \sum_{i} \mu_{i} \, \sigma_{i, SM} \times A_{if}^{k} \times \epsilon_{if}^{k} \right) \times \mu_{f} \times B_{f, SM} \times \mathcal{L}^{k}
\end{equation}

where $A$ represents the detector acceptance, $\epsilon$ the reconstruction efficiency and $\mathcal{L}$ the integrated luminosity.
The test statistic used is of the type

\begin{equation}
\Lambda(\mu) = \frac{L(\mu, \hat{\hat{\theta}}(\mu))}{L(\hat{\mu}, \hat{\theta})}
\end{equation}

where $\mu$ is the parameter of interest, $\theta$ the nuisance parameters, $L(\hat{\mu}, \hat{\theta})$ the global likelihood maximum, and $L(\mu, \hat{\hat{\theta}})$ likelihood maximum for tested $\mu$ point.

\section{\label{sec:sm_analyses} Standard Model analyses}
SM analyses were performed for various possible decay modes. The observation of a new boson is well established in the two-photon channel \cite{ATLAS-CONF-2013-012}, in the golden four-lepton decay channel \cite{ATLAS-CONF-2013-013} as well as in the $WW$ channel \cite{ATLAS-CONF-2013-030}. Searches were performed for the bosonic $H \rightarrow Z\gamma$ channel and the search for higher mass states in the $H \rightarrow ZZ \rightarrow 4l$ channel was updated. The corresponding integrated luminosities for the datasets used for these analysis are $4.8 \, \mathrm{fb}^{-1}$ for the data collected at $\sqrt{s} = 7 \, \mathrm{TeV}$ and $20.7 \, \mathrm{fb}^{-1}$ for the data collected at $\sqrt{s} = 8 \, \mathrm{TeV}$. Searches were performed for a SM Higgs Boson in the fermionic $b\overline{b}$ \cite{ATLAS-CONF-2012-161, ATLAS-CONF-2012-135} and $\tau\tau$ \cite{ATLAS-CONF-2012-160} channels, which however are not yet conclusive. In this case the dataset at $\sqrt{s} = 8 \, \mathrm{TeV}$ is limited to a corresponding integrated luminosity of $13.0 \, \mathrm{fb}^{-1} $.
Combinations of the results were performed for the mass and signal strengths measurement  \cite{ATLAS-CONF-2013-014}, and the coupling measurements \cite{ATLAS-CONF-2013-034}.

\subsection{\label{sec:gamma_gamma}Two-photon channel}

\begin{figure}[h]
\includegraphics[scale=.4]{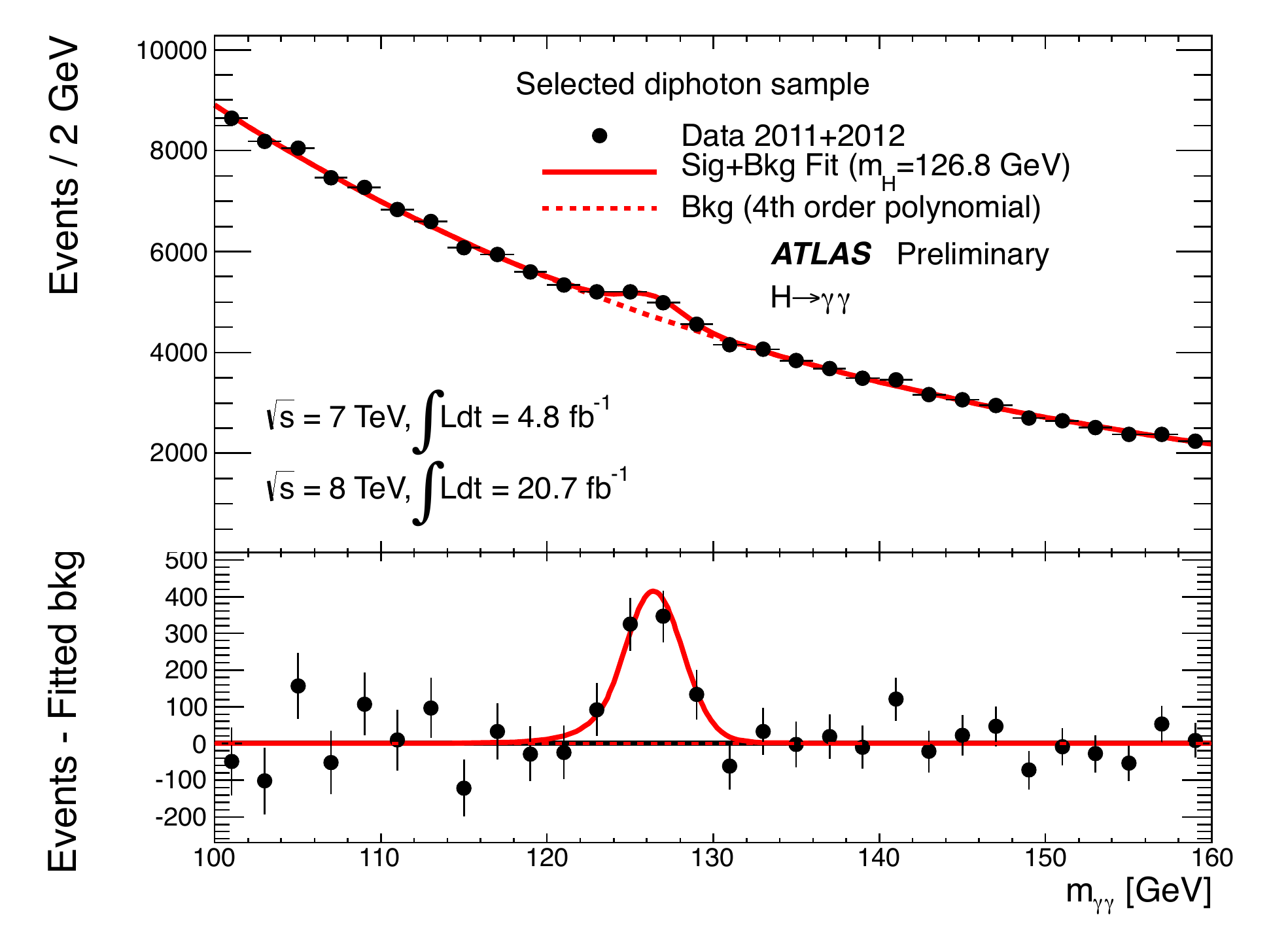}
\caption{\label{fig:gamma_gamma_fit} Invariant mass distribution of diphoton candidates for the combined $\sqrt{s}$ = 7 TeV and $\sqrt{s}$ = 8 TeV data samples. The result of a fit to the data of the sum of a signal component fixed to $m_{H} = 126.8\,  \mathrm{GeV}$ and a background component described by a fourth-order Bernstein polynomial is superimposed \cite{ATLAS-CONF-2013-012}.}
\end{figure}

The branching ratio of the Higgs decay to two photons is expected to be very small, about 0.2\%. This is due to the fact that the decay, as well as the main production mode, proceeds through loops. This latter fact on the other hand makes this channel sensitive to possible contributions from BSM heavy particles, which may circulate in the loops. Despite the very low branching ratio the sensitivity in this channel is very good, with an expected local significance for a 125 GeV SM Higgs boson of $4.1 \, \sigma$. This is due first of all to the very clear signature given by two back to back photons, which can be reconstructed into a resonance. The two-photon mass resolution is very good ($\sigma_{m} \sim 1.7 \, \mathrm{GeV}$) and was found to be stable against time and pileup conditions. The mass resolution was improved thanks to the negligible uncertainty on the primary vertex identification achieved via calorimeter pointing. 
The analysis was performed in 14 categories targeting different production modes: VH, where additional requirements on leptons, jets and $E_{T}^{\mathrm{miss}}$ in the event were applied, and two VBF categories.
The main backgrounds in this channel are given by irreducible $\gamma \gamma$ continuum (75\%) and $\gamma$-jet and jet-jet events (25\%). The two latter backgrounds are reduced by means of tight photon identification and isolation requirements. 
For each category the background was parametrised by an analytic function. The model was chosen using Monte-Carlo (MC) simulated samples in order to minimise biases.
The background was extrapolated in the signal region mass window from the side-bands in data and the signal to background ratio is of order 3\%. The inclusive fit to data of a background plus signal model is shown in Figure \ref{fig:gamma_gamma_fit}.

The observed local significance is  $7.4 \, \sigma$ ($4.1 \, \sigma$ SM expected).  
 
\begin{figure}[t]
\includegraphics[scale=.4]{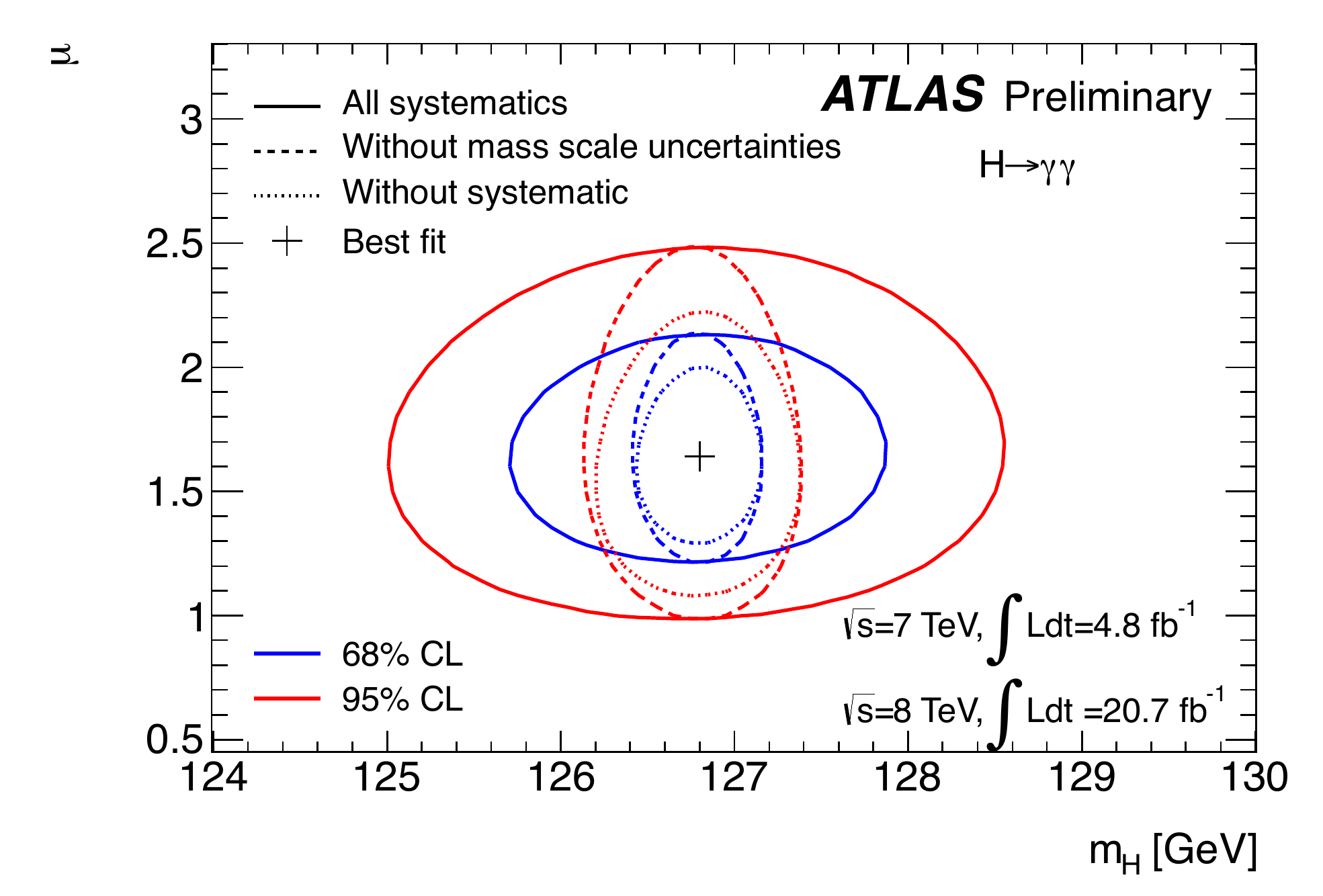}
\caption{\label{fig:gamma_gamma_mass_mu} The best-fit values of $m_{H}$ and $\mu$ for the two-photon channel and their 68\% (blue) and 95\% (red) CL contours. Results when photon energy scale systematic uncertainties are removed (dashed), and results when all systematic uncertainties are removed (dotted), are also shown \cite{ATLAS-CONF-2013-012}.}
\end{figure}

The best-fit mass is $m_{H} = 126.8 \, \pm 0.2 \, \mathrm{(stat)} \, \pm 0.7 \, \mathrm{(syst)} \, \mathrm{GeV}$. The main systematics on the mass measurement come from the extrapolation of the photon energy scale from $Z \rightarrow ee$ analysis (0.3\%), the modelling of material in front of the calorimeter (0.3\%) and to the calorimeter presampler energy scale (0.1\%).
The signal strength at the best-fit mass is $\mu = 1.65 \, \pm \, 0.24 \, \mathrm{(stat)} ^{+0.25}_{-0.18} \, \mathrm{(syst)}$. 
The best-fit values of $m_{H}$ and $\mu$, with 68\% and 95\% confidence level (CL) contours are shown in Figure \ref{fig:gamma_gamma_mass_mu}.
A fiducial cross-section is defined for the kinematic range  $E_{T}^{\gamma^{1}} > 40 \, \mathrm{GeV}, E_{T}^{\gamma^{1}} > 30 \, \mathrm{GeV}$ and $|\eta^{\gamma}| < 2.37$, where $E_{T}^{\gamma^{1,2}}$ are the leading and subleading photon transverse energies, and $\eta^{\gamma}$ is the photon pseudorapidity. 
The inclusive fiducial cross-section is measured to be $\sigma_{\mathrm{fid}} \times BR = 56.2 \pm 12.5 \, \mathrm{fb}$.

\begin{figure}[t]
\includegraphics[scale=.4]{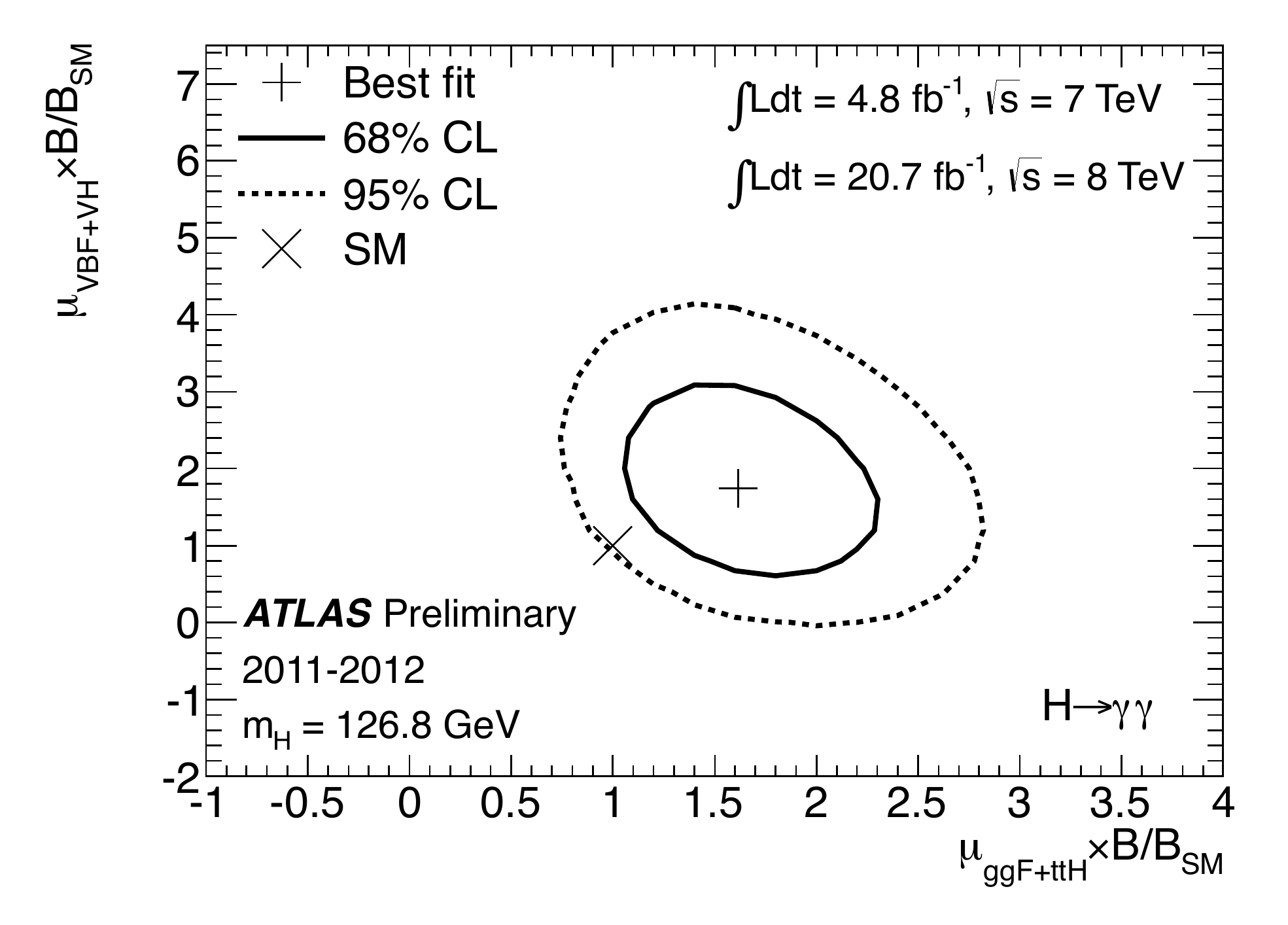}
\caption{\label{fig:gamma_gamma_couplings} The best-fit values ($+$) of $\mu_{\mathrm{ggF+t\overline{t}H}} \times B/B(SM)$ and $\mu_{\mathrm{VBF+VH}} \times B/B(SM)$ for the two-photon channel and their 68\% (solid) and 95\% (dashed) CL contours. The expectation for a SM Higgs boson is also shown ($\times$) \cite{ATLAS-CONF-2013-012}.}
\end{figure}

Exploiting the fact categories target different production modes it was possible to obtain information on the couplings. Separate signal strengths for the $\mathrm{ggF+t\overline{t}H}$ production modes (sensitive especially to the couplings to fermions) and VBF+VH (sensitive to couplings to bosons) were fit. The best-fit values of $\mu_{\mathrm{ggF+t\overline{t}H}} \times B/B_{SM}$ and $\mu_{\mathrm{VBF+VH}} \times B/B_{SM}$ are shown in Figure \ref{fig:gamma_gamma_couplings} and are in agreement with the SM expectations.

\begin{figure}[h]
\includegraphics[scale=.2]{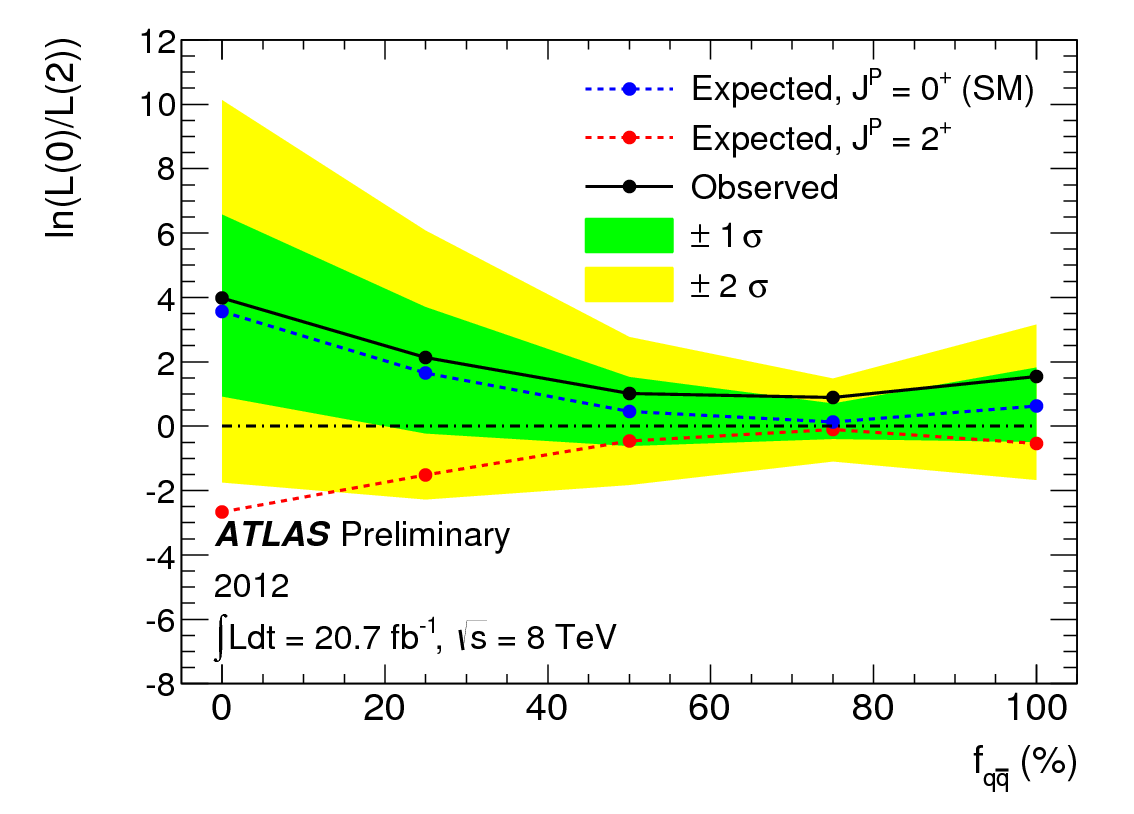}
\caption{\label{fig:gamma_gamma_spin_limit} The expected log-likelihood ratio for the null $J^{P} = 0^{+}$ (blue) and alternative $J^{P} = 2^{+}$ (red) hypotheses as well as the observed values as a function of the $q\overline{q}$ fraction of the spin-2 signal production in the two-photon channel \cite{ATLAS-CONF-2013-029}.}
\end{figure}

The spin-parity properties of the new boson were analysed by studying the $dN/d|\cos\theta^{*}|$, where $\theta^{*}$ is the polar angle of the photons with respect to the z-axis of the Collins-Soper frame. This distribution is expected to be flat before applying any cuts for $J^{P} = 0^{+}$, as predicted for the SM Higgs boson. The Landau-Yang theorem forbids the decay of spin-1 particle into a pair of photons. It is
therefore strongly disfavored by the observation in the two-photon channel. The possibility that the new particle has $J^{P} = 2^{+}$ was considered, using a spin-2 graviton-like model with minimal couplings. The spin-2 resonance can be produced either via gluon fusion or via P-wave quark-antiquark annihilation ($q\overline{q}$). Five scenarios corresponding to different admixtures of the production modes were considered.  For the gluon fusion production mode the angular distribution is given by $dN/d|\cos\theta^{*}| = 1 + 6 \cos^{2}{\theta^{*}} + \cos^{4}{\theta^{*}}$, while for the quark-antiquark annihilation it follows $dN/d|\cos\theta^{*}| = 1 - \cos^{4}{\theta^{*}}$. The expected log-likelihood ratio for the null $J^{P} = 0^{+}$ and alternative $J^{P} = 2^{+}$ hypotheses, as well as the observed values as a function of the $q\overline{q}$ fraction of the spin-2 signal production, are shown in Figure \ref{fig:gamma_gamma_spin_limit}.

\subsection{\label{sec:gamma_gamma}Four-lepton channel}
\begin{figure}[t]
\includegraphics[scale=.32]{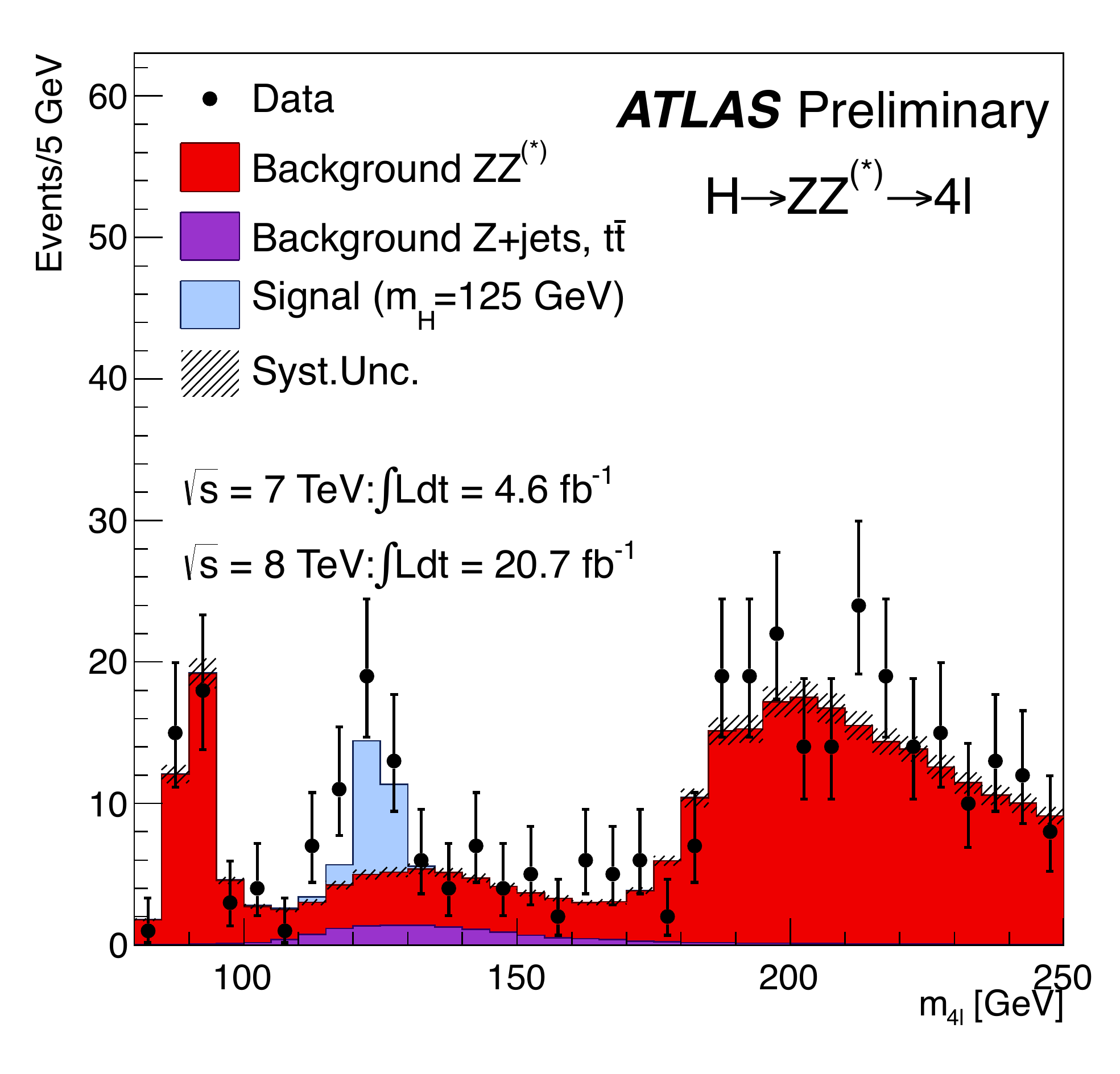}
\caption{\label{fig:4leptons_mass} The distribution of the four-lepton invariant mass, $m_{4l}$, for the selected candidates in the four-lepton channel compared to the background expectation for the combined $\sqrt{s}= 8 \, \mathrm{TeV}$ and $\sqrt{s}=7 \, \mathrm{TeV}$ datasets. The signal expectation for the $m_{H}=125 \, \mathrm{GeV}$ hypothesis is also shown \cite{ATLAS-CONF-2013-013}.}
\end{figure}

The channel where the Higgs decays to four leptons, $H \rightarrow ZZ^{*} \rightarrow l^{+}l^{-}l^{+}l^{-}$ is the golden channel for $m_{H} = 125 \, \mathrm{GeV}$. Despite the low decay branching ratio, this channel has in fact a very clear signature and it is particularly clean from the background. Moreover it can benefit from a very good mass resolution.
The real challenge of this analysis is to maximise the acceptance, which was achieved thanks to the high reconstruction and identification efficiencies of electrons and muons to transverse momentum thresholds $p_{T} > 7/6 \, \mathrm{GeV}$ respectively. Furthermore, the recovery of final state radiation photons allowed to increase the signal acceptance by $4\%$.
The mass resolution was improved by means of a kinematic fit including a Z mass constraint. 
The residual background is mostly due to irreducible $ZZ$ continuum
which was estimated from MC simulation. Contributions from $Z + \mathrm{jets}$, $Z \rightarrow b\overline{b}$ and $t\overline{t}$ pairs were reduced by means of lepton isolation and impact parameter cuts and were estimated making use of data control regions. 
The breakdown of the background prediction for the full data sample considered is shown in Figure \ref{fig:4leptons_mass}, where the contribution from a SM Higgs with $m_{H} = 125 \, \mathrm{GeV}$ is shown and data are superimposed as well.

\begin{figure}[t]
\includegraphics[scale=.32]{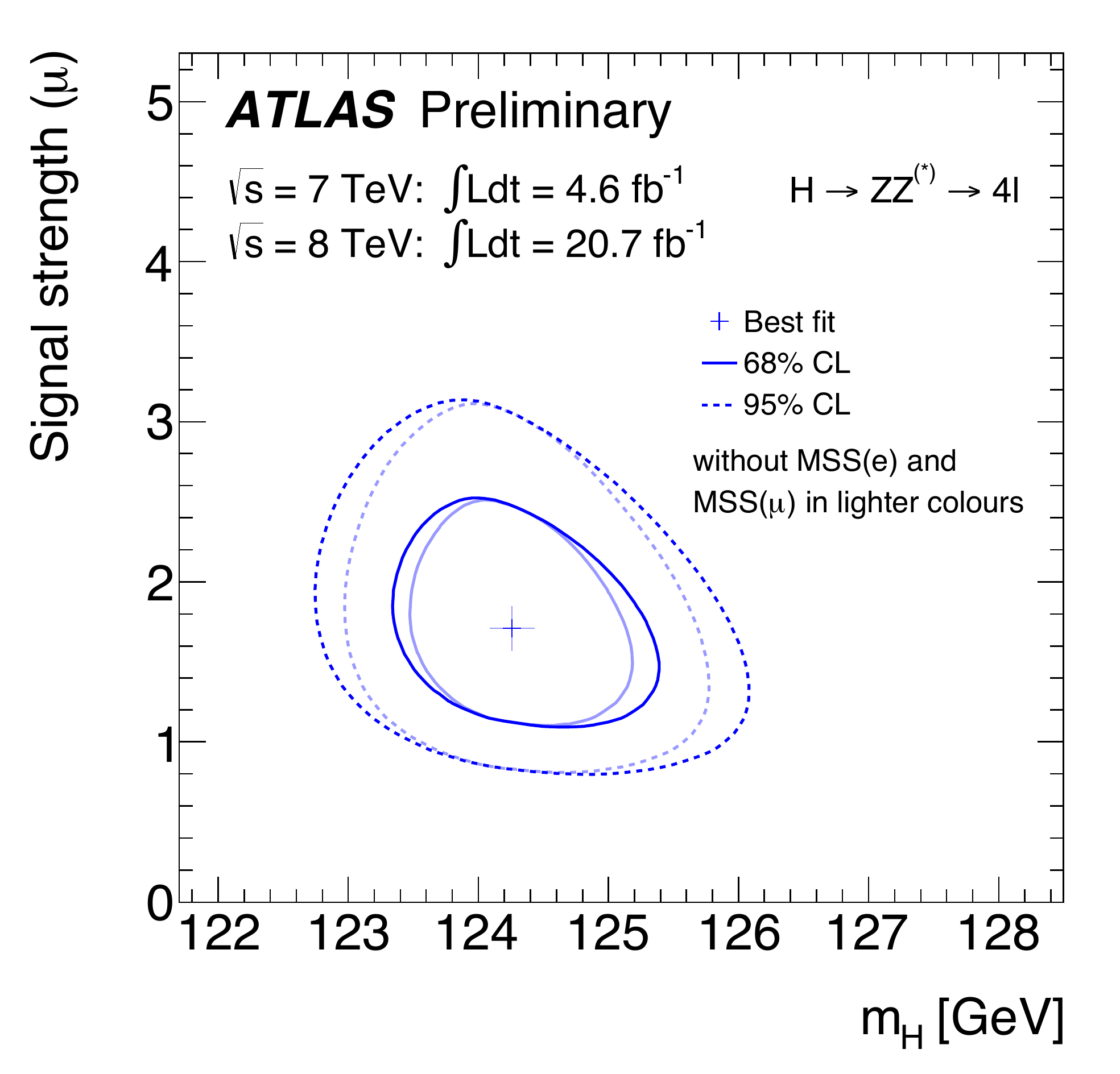}
\caption{\label{fig:4leptons_mass_mu} Likelihood ratio contours in the $(\mu, m_{H})$ plane that, in the asymptotic limit, correspond to 68\% and 95\% level contours, shown with (dark colour curves) and without mass scale uncertainties (MSS(e) and MSS(mu)) applied (lighter colour curves) for the four-lepton channel \cite{ATLAS-CONF-2013-013}.}
\end{figure}

The local significance of the excess visible in the data is $6.6 \, \sigma$ at a best measured mass of $m_{H} = 124.3^{+0.6}_{-0.5} \, \mathrm{(stat)}^{+0.5}_{-0.3} \, \mathrm{(syst) \, GeV}$. 
The systematic uncertainty on the mass measurement is dominated by contributions related to the muon momentum measurement, since the $4\mu$ channel is the one with largest yield. For the channels involving electrons similar systematic uncertainties as those in the two-photon channel apply.
The signal strength at this mass is $\mu = 1.7^{+0.5}_{-0.4}$.
The best-fit values of $m_{H}$ and $\mu$, with 68\% and 95\% CL contours are shown in Figure \ref{fig:4leptons_mass_mu}.

\begin{figure}[t]
\includegraphics[scale=.32]{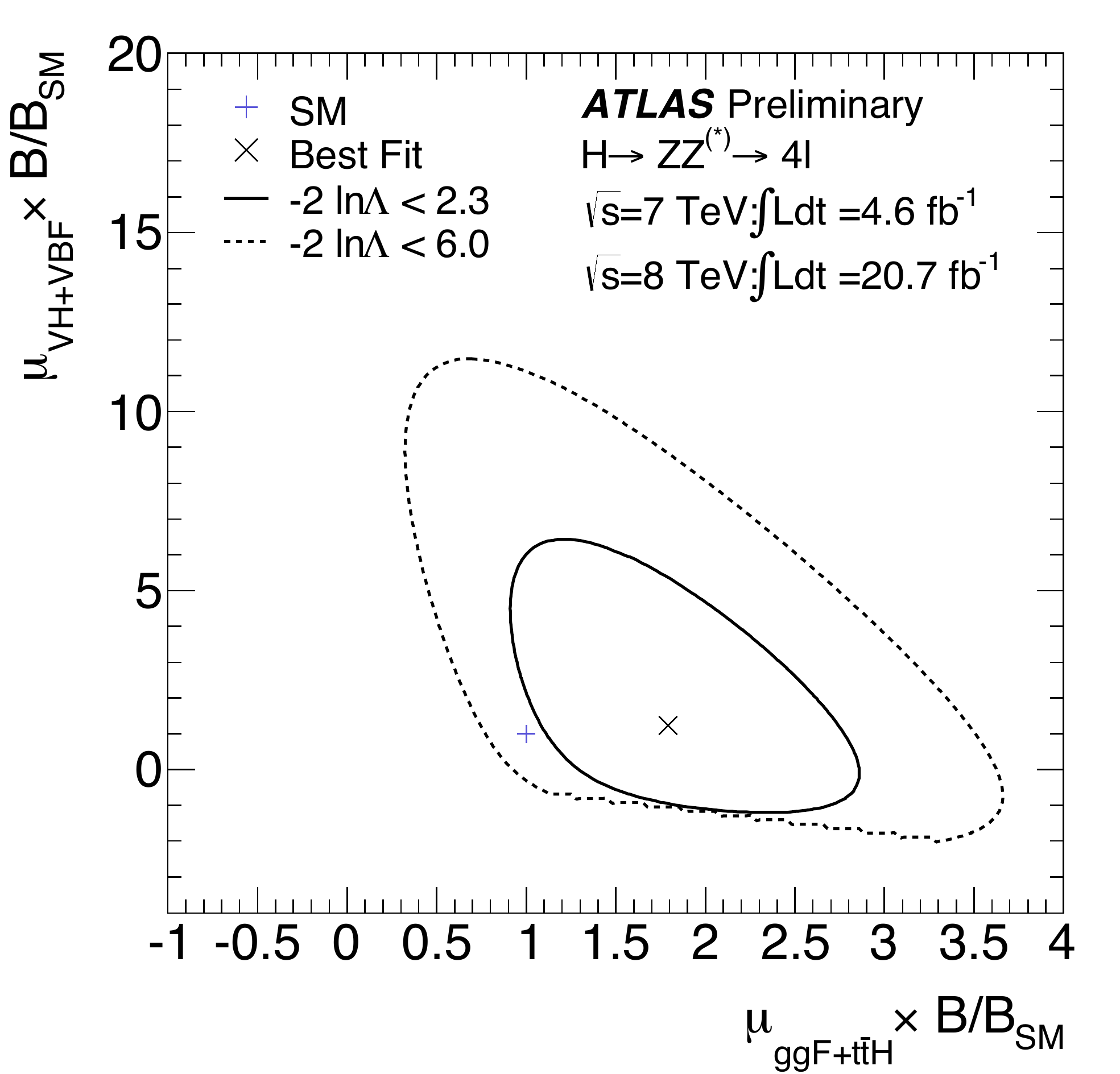}
\caption{\label{fig:4leptons_couplings}Likelihood contours in the $(\mu_{\mathrm{ggF+t\overline{t}H}}, \, \mu_{\mathrm{VBF+VH}})$ plane including the branching ratio factor B/B(SM) for the four-lepton channel. The best-fit to the data ($\times$) and $- \ln{\Lambda} < 2.3$ (full) and $6.0$ (dashed) contours are also indicated, as well as the SM expectation ($+$)  \cite{ATLAS-CONF-2013-013}.}
\end{figure}

Three event categories were used in the analysis, targeting different production modes. A VBF category was defined by requiring two high transverse momentum jets, with a pseudorapidity separation $\Delta \eta_{jj} > 3$ and an invariant mass $m_{jj} > 350 \, \mathrm{GeV}$. A VH category was obtained requiring the presence of an additional lepton. Events that do not belong to the VBF or VH categories result in a sample enriched in ggF events. 
The categorisation allowed to obtain information on the couplings of the observed boson, similarly to what was done for the two-photon channel. Figure \ref{fig:4leptons_couplings} shows the best-fit values of $\mu_{\mathrm{ggF+t\overline{t}H}} \times B/B_{SM}$ and $\mu_{\mathrm{VBF+VH}} \times B/B_{SM}$, which are in agreement at the $< 2\sigma$ level with the SM expectation. 

\begin{figure}[h]
\includegraphics[scale=.17]{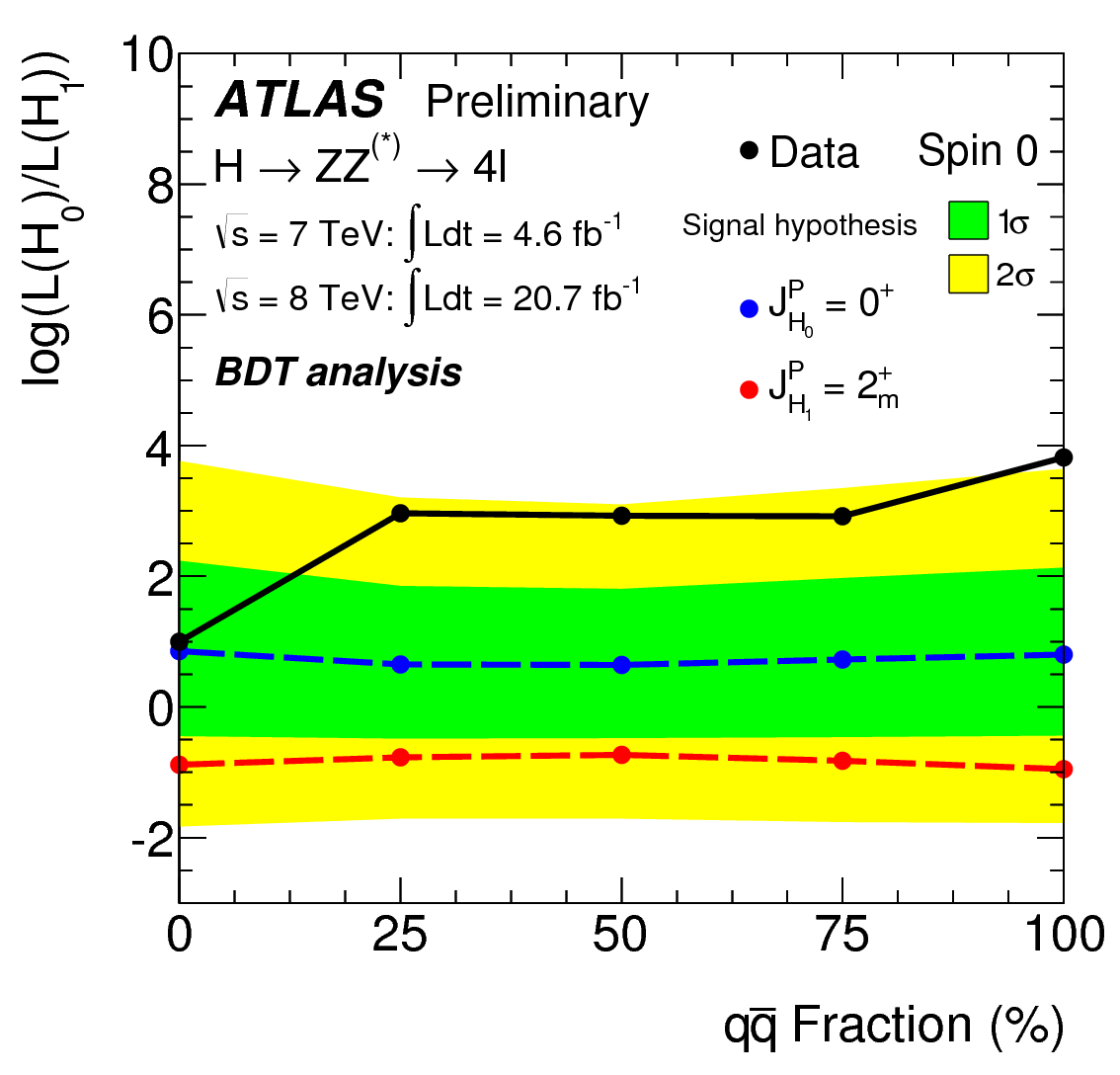}
\caption{\label{fig:4leptons_fig_18a} Variation of the medians of the log-likelihood ratio distribution generated for varying fractions of $q\overline{q}$ in a mixed $q\overline{q}$ and ggF production for testing the $2^{+}$ hypothesis when assuming the $0^{+}$ hypothesis in the four-lepton channel. The blue and red data points correspond to the median values for the $0^{+}$ and $2^{+}$ hypotheses, respectively, for each fraction. The black points represent the log-likelihood values observed in data. \cite{ATLAS-CONF-2013-013}.}
\end{figure}

A spin-parity analysis was performed on 43 events with reconstructed four-lepton invariant mass $m_{4l}$ satisfying $115 \, \mathrm{GeV} < m_{4l} < 130 \, \mathrm{GeV}$.
Six hypotheses for spin-parity states were tested, namely $J^{P} \; 0^{+}, 0^{-}, 1^{+}, 1^{-}, 2^{+}, 2^{-}$. The spin-2 states correspond to a graviton-like tensor with minimal couplings. As for the two-photon channel case different gluon fusion and $q\overline{q}$ production admixtures were considered.
Discriminants were built from seven spin sensitive observables: the two $Z$ masses, 1 production and  4 decay angles. Both an analysis using Boosted Decision Trees and one employing a matrix element based likelihood ratio (MELA) were developed. 
The $0^{+}$ hypothesis is favoured, and in particular the $0^{-}$ and $1^{+}$ hypotheses are excluded at $> 97\% \, \mathrm{CL}$.
Figure \ref{fig:4leptons_fig_18a} shows the results for the $0^{+}$ versus $2^{+}$ hypotheses for varying  $q\overline{q}$ production fraction.

\subsection{\label{sec:ww}$WW^{*} \rightarrow l\nu l\nu$  channel}
\begin{figure}[b]
\includegraphics[scale=.17]{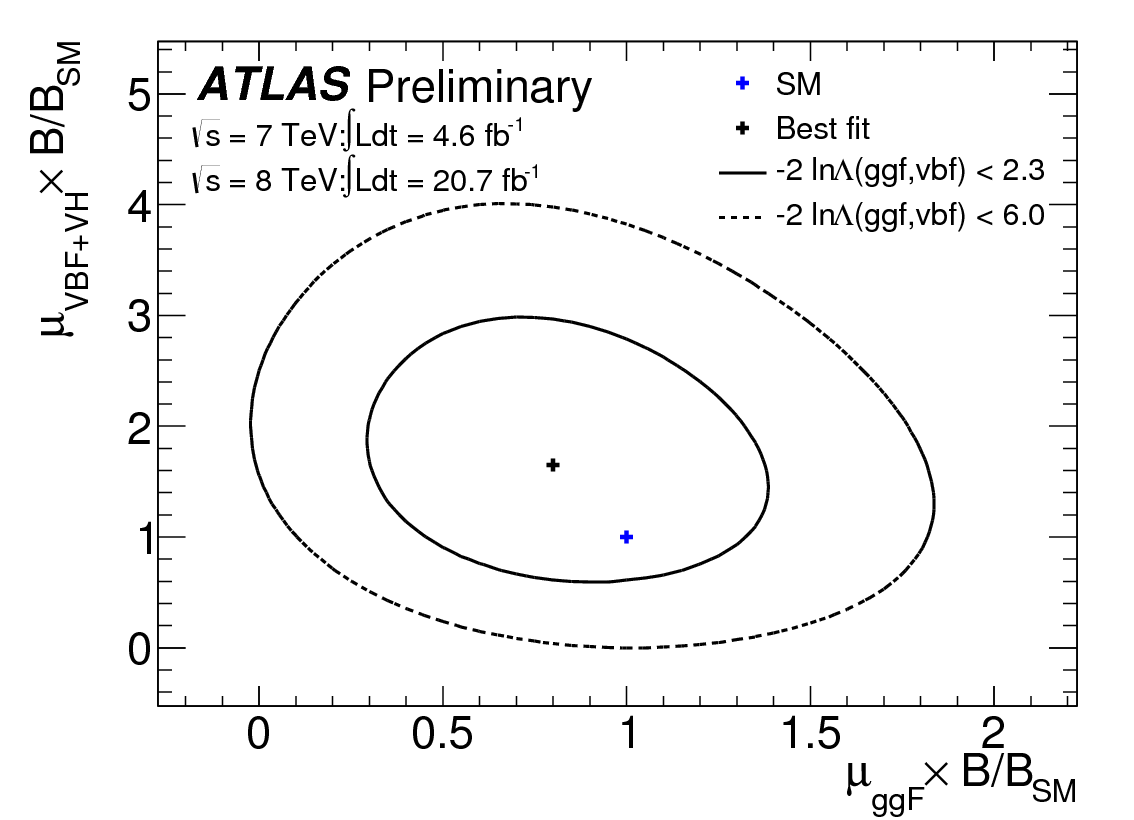}
\caption{\label{fig:ww_couplings_counturs}Likelihood contours for separate ggF and VBF signal strength parameters in the $WW$ channel \cite{ATLAS-CONF-2013-030}.}
\end{figure}

At $m_{H} = 125 \, \mathrm{GeV}$ the branching ratio for this channel is quite large, $\sim 20\%$, despite being below the real $WW$ decay threshold. This channel has a clear signature given by the presence of two charged leptons and missing transverse momentum due to the neutrinos emitted in the $W$ decays. However it is experimentally difficult due to the fact full mass reconstruction is not possible. The main backgrounds come from irreducible WW continuum, $t\overline{t}$ pairs and $W \rightarrow l \nu$, which are estimated making use of data control regions. 
An excess is observed in the data, which at $m_{H} = 125 \, \mathrm{GeV}$ has a local significance of $3.8 \, \sigma$ (expected $3.7 \, \sigma$). This corresponds to a signal strength $\mu = 1.0 \pm 0.3$. 
Even in this case separate signal strengths were fitted for different production modes. Figure \ref{fig:ww_couplings_counturs} shows the likelihood contours for the ggF and VBF signal strengths. 
The signiÞcance of the excess in VBF categories is $2.5 \, \sigma$ (expected $1.6 \, \sigma$).

The measured value of the product of the cross-section and the $WW^{*}$ branching ratio for a signal at $m_{H} = 125 \, \mathrm{GeV}$ at $8 \, \mathrm{TeV}$ is $6.0 \pm 1.6 \, \mathrm{pb}$ while the expected value is $4.8 \pm 0.7 \, \mathrm{pb}$.

\begin{figure}[h]
\includegraphics[scale=.17]{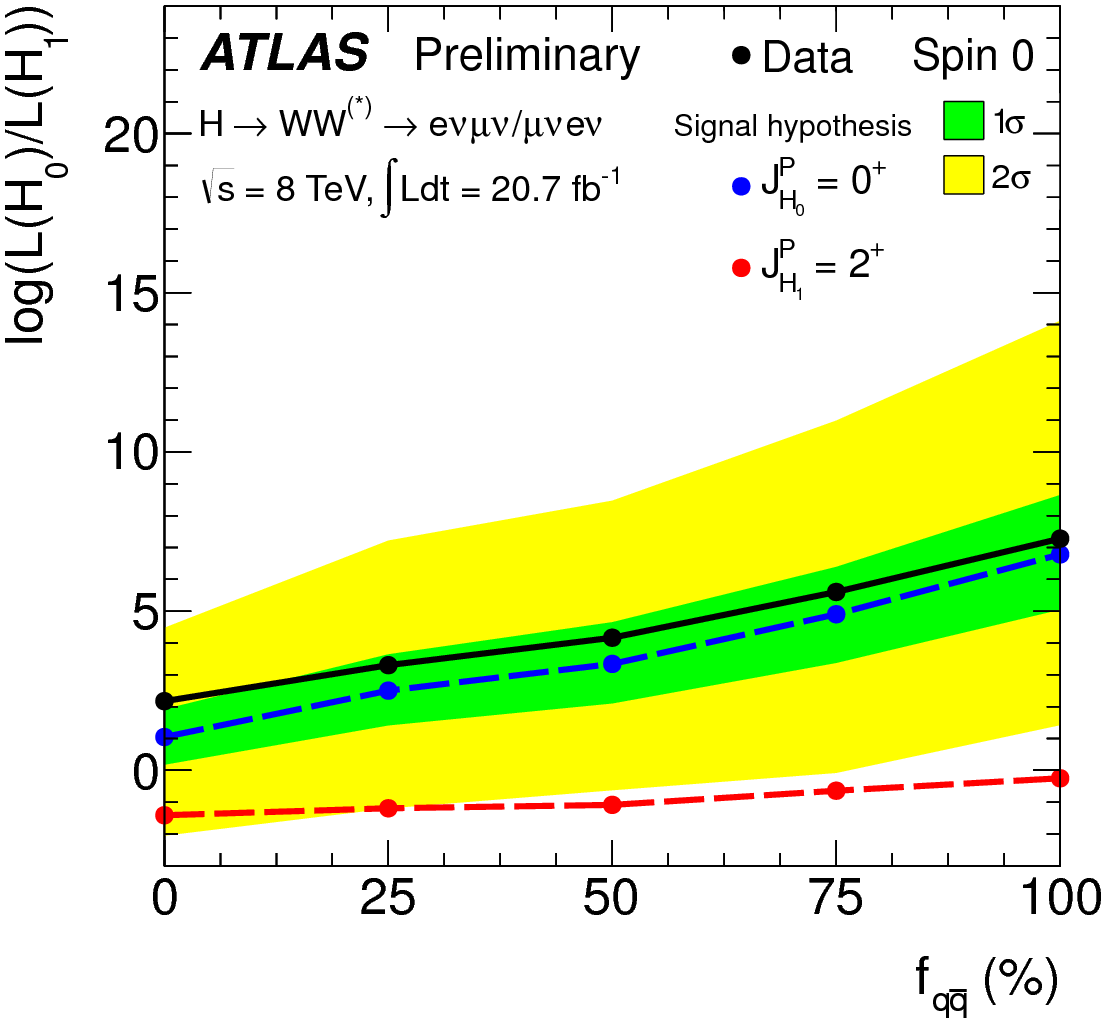}
\caption{\label{fig:ww_spin_new}The median test statistic for $0^{+}$ (blue dashed line) and $2^{+}$ (red dashed line) as well as the observed value (solid black line), for various assumptions of  $f_{q\overline{q}}$, of a $2^{+}$ particle for the $WW$ channel. The $\pm \, 1 \sigma$ and $\pm \, 2 \sigma$ uncertainty bands for $0^{+}$ are also shown by the green and yellow regions, respectively \cite{ATLAS-CONF-2013-031}.}
\end{figure}

The spin-parity properties of the observed particles were tested in this case as well. Angular distributions were exploited, the main variables being the dilepton mass $m_{ll}$ and the azimuthal angular distance $\Delta \phi_{ll}$ of the two leptons. Only the different flavour, zero-jet channels are used for the spin measurement, since they give the best sensitivity.
The $0^{+}$ and $2^{+}$ hypotheses were tested, finding $0^{+}$ favoured against $2^{+}$ at $ 95\% \, \mathrm{CL}$.
Figure \ref{fig:ww_spin_new} shows the results of the spin-parity analysis as a function of the $q\overline{q}$ production fraction$f_{q\overline{q}}$ for the graviton-like tensor with minimal couplings model used for the $2^{+}$ hypothesis.
It should be noted that this analysis is sensitive in the $f_{q\overline{q}}$ region where the two-photon spin analysis looses discrimination power.

\subsection{\label{sec:Z_gamma}$Z\gamma$ channel}
\begin{figure}[h]
\includegraphics[scale=.4]{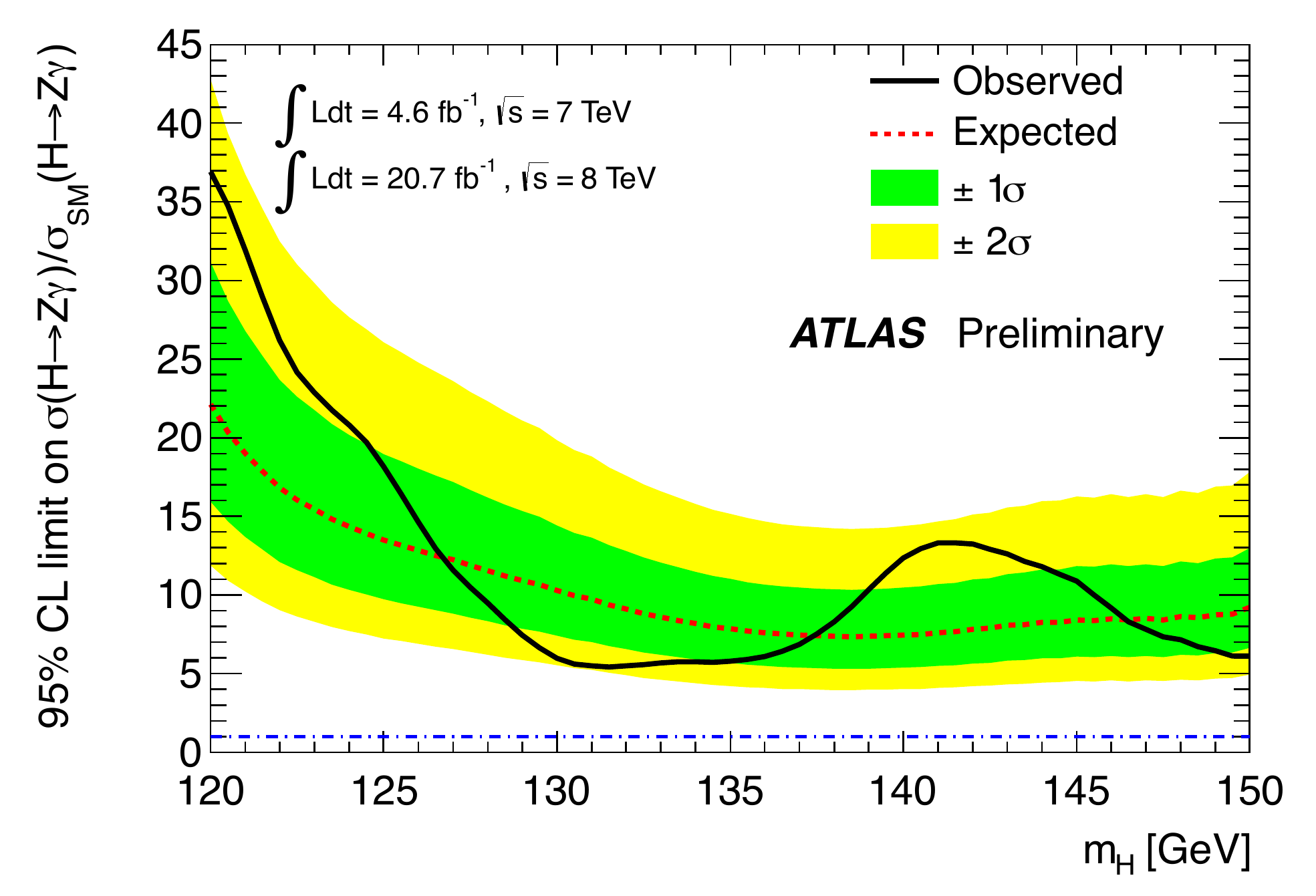}
\caption{\label{fig:Zgamma_limit}Observed 95\% CL limits (solid black line) on the production cross-section of a SM Higgs boson decaying to $Z \gamma$, as a function of the Higgs boson mass. The median expected 95\% CL exclusion limits (dashed red line) are also shown. The green and yellow bands correspond to the $\pm \, 1\sigma$ and $\pm \, 2 \sigma$ intervals \cite{ATLAS-CONF-2013-009}.}
\end{figure}

The $H \rightarrow Z \gamma$ process is very rare, with a branching ratio at $m_{H} = 125 \, \mathrm{GeV}$ $B(H \rightarrow Z\gamma) = 1.54 \times 10^{-3}$, but particularly interesting since it proceeds through loops, and could therefore be sensitive to new physics.
The analysis was performed with $Z \rightarrow ee$ and  $Z \rightarrow \mu\mu$. 
The main backgrounds are irreducible $Z + \gamma$, where the photon is produced through either initial- or final-state radiation, and a reducible contribution from $Z+\mathrm{jets}$.
The discriminating variable considered is the difference between the three body and dilepton mass $\Delta m = m_{ll \gamma} - m_{ll}$. This variable was chosen because it is almost independent of lepton energy scales and it is to a large extent insensitive to the contribution to the signal from final state radiation in $H \rightarrow \mu\mu$ decays.
The background was estimated following an approach similar to the two-photon analysis. Side-bands fits were used to extrapolate the background in the signal region and the estimate was cross-checked with data-driven methods.
Since no excess over the expected background was observed, limits were set and are shown in Figure \ref{fig:Zgamma_limit}. The limit at $m_{H} = 125 \, \mathrm{GeV}$ is $18.2 \times \sigma_{SM}$ (expected $13.5 \times \sigma_{SM}$).

\subsection{\label{sec:diboson_high_mass}Other Higgs to bosons channels: search for higher mass states.}
\begin{figure}[t]
   \subfigure[]{ 
        \begin{overpic}[width=0.22\textwidth]{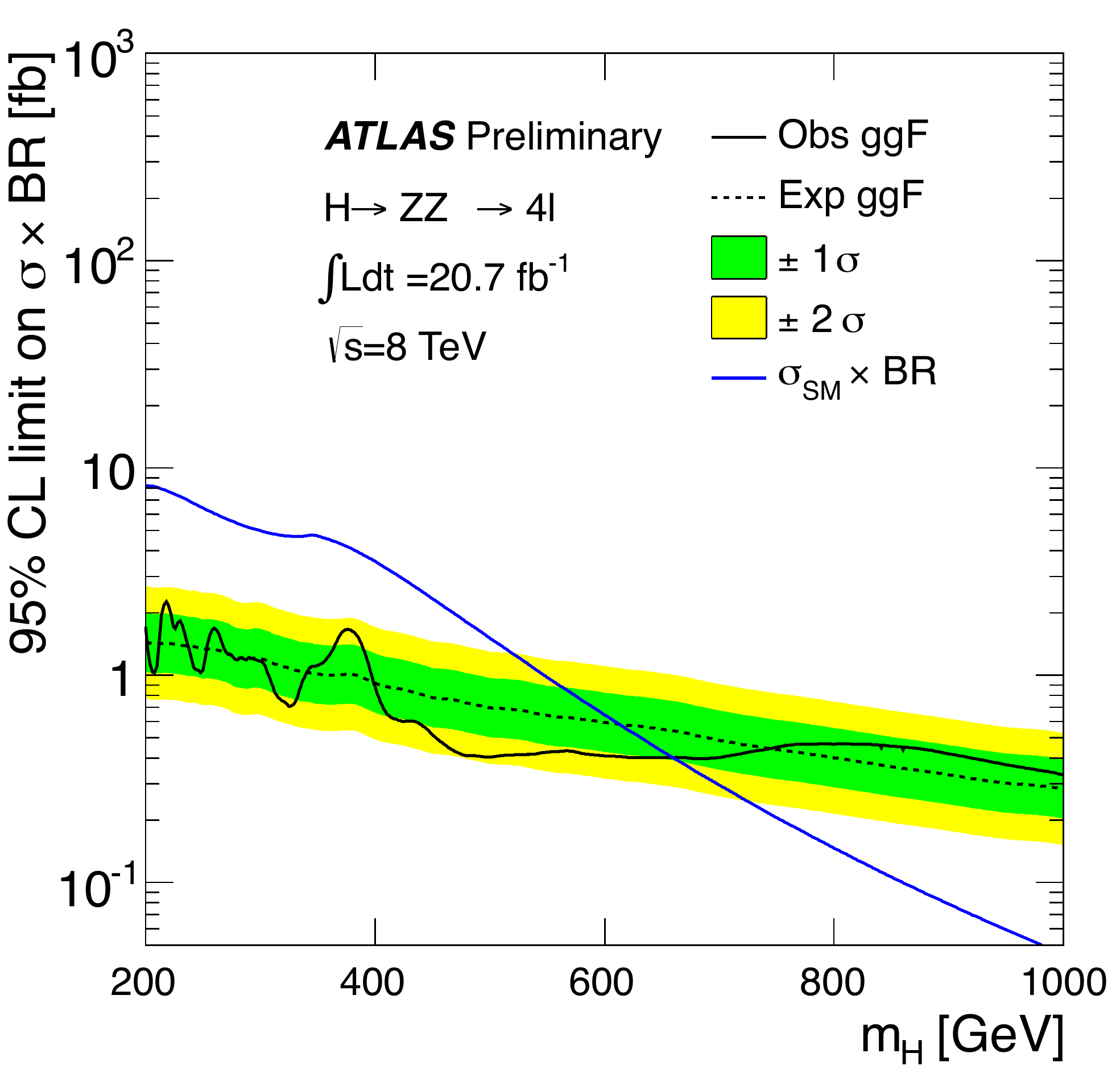}
        \end{overpic}
    }  
     \subfigure[]{ 
        \begin{overpic}[width=0.22\textwidth]{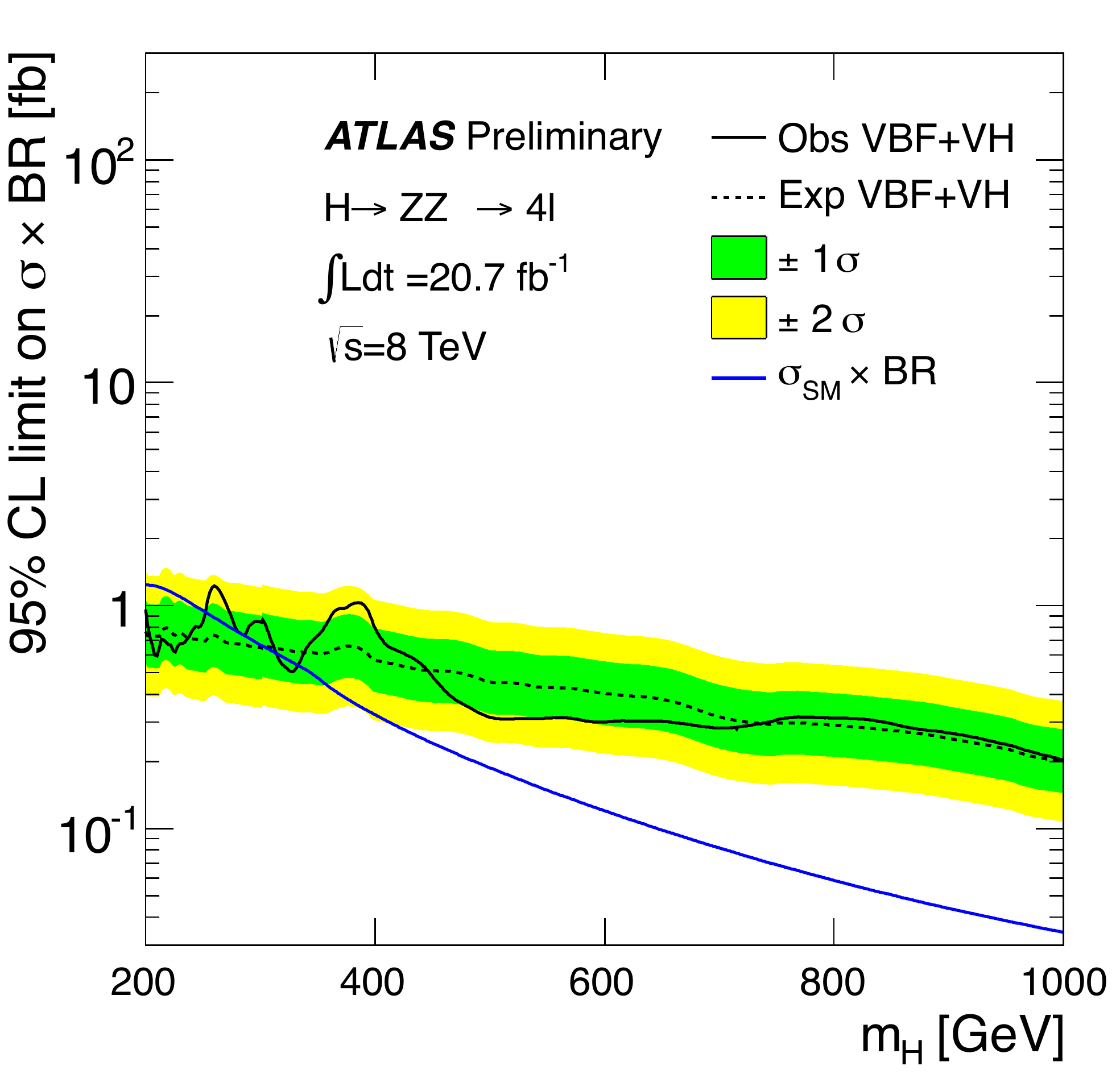}
        \end{overpic}
    }  
\caption{\label{fig:4leptons_highmass}Expected and observed 95\% CL upper limit on the production cross-section times branching ratio of $H \rightarrow ZZ  \rightarrow 4l$ ($l=e,\mu$) for an (a) ggF and (b) VBF/VH produced SM-like signal as a function of $m_{H}$ \cite{ATLAS-CONF-2013-013}.}
\end{figure}

The SM Higgs boson is excluded in the mass range $111 - 559 \, \mathrm{GeV}$ with the exception of the excess at $m_{H} \sim 125 \, \mathrm{GeV}$ \cite{Aad20121}. The sensitivity in the high mass region was achieved thanks to the $H \rightarrow WW$ and $H \rightarrow ZZ$ analyses based on the dataset collected at $\sqrt{s} = 7\,  \mathrm{TeV}$. It is now interesting to revisit this kind of analyses in the perspective of the discovery of a new boson, especially to search for further possible states at higher mass. The $H \rightarrow ZZ \rightarrow 4l$ was updated to the full LHC datasets. 
In the analysis the signal was assumed to have a SM-like width, estimated using the complex-pole-scheme. 
No excess was observed above the background estimation. Limits were set on the cross-section times the $H \rightarrow ZZ$ branching ratio for the ggF and VBF production processes separately to allow for different rates for the two production modes. The results are shown in Figure \ref{fig:4leptons_highmass}.

\clearpage

\subsection{\label{sec:tautau} $\tau \tau$ channel}
\begin{figure}[h]
\includegraphics[scale=.15]{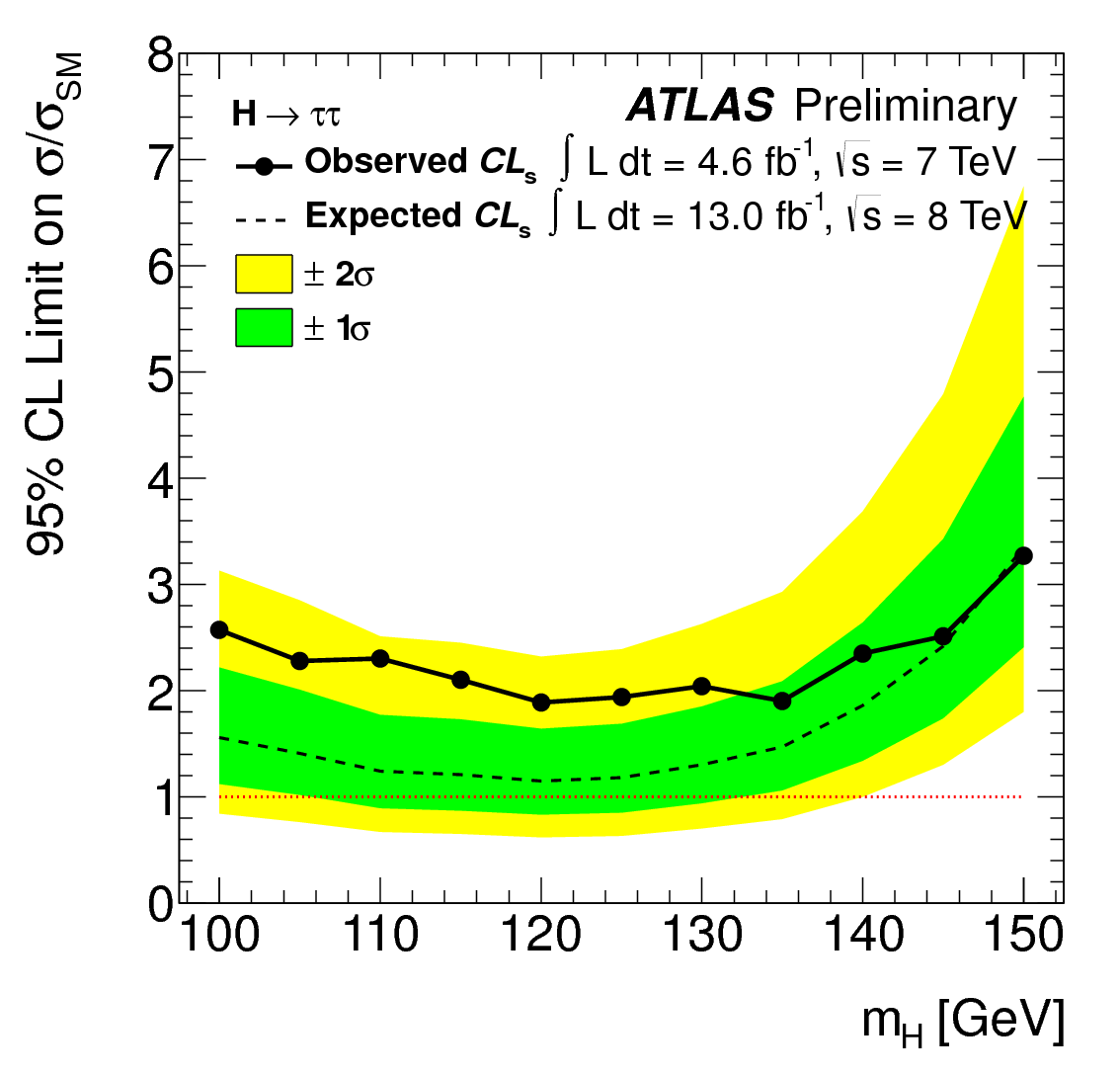}
\caption{\label{fig:tautau_limit_combined_2011_2012}Observed (solid) and expected (dashed) 95\% confidence level upper limits on the Higgs boson cross-section times branching ratio for the decay to two taus, normalised to the SM expectation, as a function of the Higgs boson mass. The bands around the dashed line indicate the $\pm 1 \, \sigma$ and $\pm \, 2 \sigma$ uncertainties of the expected limit \cite{ATLAS-CONF-2012-160}.}
\end{figure}

The search for fermionic decays of the newly discovered resonance is of uttermost importance to understand its properties. The channel $H \rightarrow \tau \tau$ is only the second in branching ratio among fermionic channels, after $H \rightarrow b \overline{b}$, but provides the largest sensitivity at $m_{H} = 125 \, \mathrm{GeV}$. 
Tau leptons can decay either hadronically or leptonically. Both types of decay were employed in the search. Hadronic tau decays were reconstructed from narrow and low track multiplicity jets and were identified against quark or gluon initiated jets by means of a BDT discriminator based on tracking and calorimeter information. 
Ten analysis categories were used, according to decay type (lepton-lepton, lepton-hadron and hadron-hadron) and jet and event topology properties. These were used in order to target different production modes.
In a hadronic environment this channel is subject to a considerable background contamination, and is therefore accessible for $m_{H}$ as low as 125 GeV only through the distinctive VBF production signature. 
The background is dominated by irreducible $Z \rightarrow \tau \tau$. This background was estimated by means of a hybrid data-MC technique. $Z \rightarrow \mu \mu$ data events were considered where muons were replaced by simulated tau leptons.
Other processes that contribute to the background are $Z + \mathrm{jets}$, $W + \mathrm{jets}$, $t\overline{t}$ pairs and single-top, which were estimated by MC normalised in data control regions. Di-boson production instead was fully estimated from MC and contributions from multi-jet events were estimated by fully data-driven techniques.
Mass reconstruction is another challenge for this analysis, since two to four neutrinos are emitted in the tau decays. The knowledge of the tau decay kinematics was used to improve the mass resolution.
No significant excess above the SM background prediction was observed, limits were therefore set that are shown in Figure \ref{fig:tautau_limit_combined_2011_2012}. The limit at $m_{H} = 125 \, \mathrm{GeV}$ is $1.9 \times \sigma_{SM}$  (expected $1.2 \times \sigma_{SM}$).

\subsection{\label{sec:tautau} $b\overline{b}$ channel}
\begin{figure}[h]
   \subfigure[]{ 
        \begin{overpic}[width=0.3\textwidth]{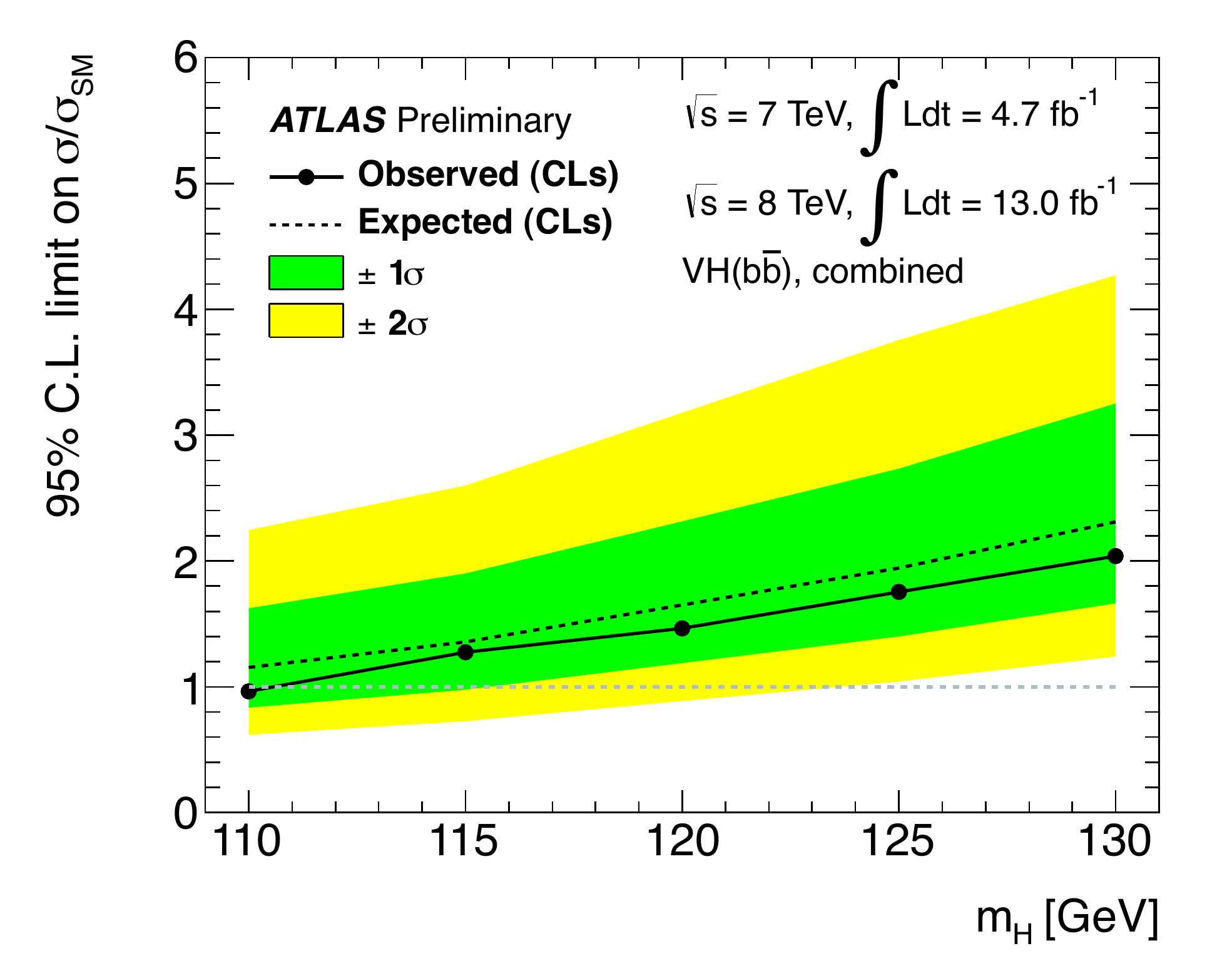}
        \label{fig:vhbb_limit}
        \end{overpic}
    }
    
      \subfigure[]{ 
        \begin{overpic}[width=0.3\textwidth]{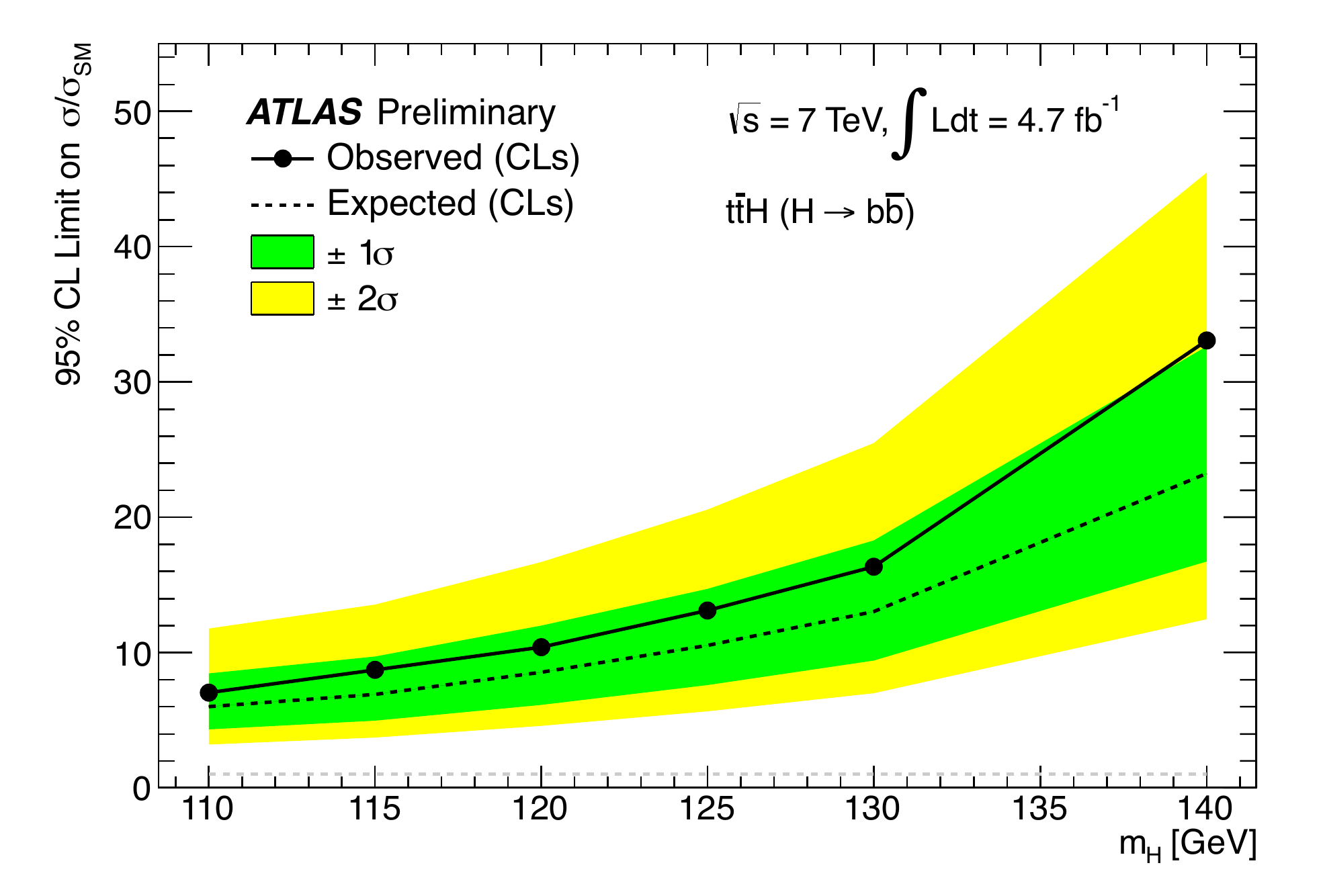}
        \label{fig:ttHbb_limit}
        \end{overpic}
    }  
    
     \caption{Expected (dashed) and observed (solid) CL limit on the normalised signal strength as a function of $m_{H}$ for the $VH, H \rightarrow b\overline{b}$ (a) and $t\overline{t}H, H \rightarrow b\overline{b}$ (b) processes \cite{ATLAS-CONF-2012-161, ATLAS-CONF-2012-135}.}

\end{figure}

The $b\overline{b}$ channel benefits from the largest decay branching ratio in the SM at $m_{H} = 125 \, \mathrm{GeV}$. However it is experimentally difficult in the hadronic environment of the LHC. Therefore specific signatures offered by the production modes need to be exploited. Two analysis were performed, one targeting the VH and the other the $\mathrm{t\overline{t}H}$ production mode.
In the associated production analysis as a baseline two b-tagged jets were required.
B-jets were tagged making use of combined information from different algorithms, based on track impact parameter significance and or secondary decay vertex reconstruction.
Three categories were considered, with zero, one or two additional leptons in the final state, targeting the $Z \rightarrow \nu \nu$, $W \rightarrow l \nu$ and $Z \rightarrow ll$ associated vector boson decay modes respectively.
Thirteen categories were then considered, which exploit different Higgs transverse boost regimes.
The mass of the $b \overline{b}$ system was reconstructed from the two b-jets in the event.
The main background contributions come from $t \overline{t}$ pairs, $W+\mathrm{jet}$ and $Z+\mathrm{jet}$.
Backgrounds shapes were taken from MC and normalised in data control regions. Exceptions are the multi-jet background, which was estimated in a fully data-driven way, and the di-boson background, which was fully derived from MC simulation.
No excess above the SM background expectation was observed. Limits were therefore set, which are shown in Figure \ref{fig:vhbb_limit}. The limit at $m_{H} = 125 \, \mathrm{GeV}$ on $\sigma_{SM}\times BR = 1.8$ (1.9 expected).
The analysis concept was cross-checked performing a $4 \, \sigma$ observation of $WZ$ production with $Z \rightarrow b\overline{b}$.

The analysis for the $\mathrm{t\overline{t}H}$ production modes requires a quite complex signature: 

\begin{equation}
t\overline{t}H \rightarrow W^{+} b \: W^{-} \overline{b} \: b\overline{b} \rightarrow l^+ \nu b \: q \overline{q}\overline{b} \: b\overline{b}
\end{equation} 

Events were selected by requiring at least 6 jets, of which at least 3 b-tagged. A kinematic likelihood fitter was used to assign objects in the detector to the objects considered in the signature above.
The main background comes from $t\overline{t}$ pair production.
No excess above the SM background expectation was observed. Limits were therefore set, which are shown in Figure \ref{fig:ttHbb_limit}. The limit at $m_{H} = 125 \, \mathrm{GeV}$ on $\sigma_{SM}\times BR = 13.1$ (10.5 expected).

\subsection{\label{sec:tautau} $\mu^{+}\mu^{-}$ channel}
\begin{figure}[h]
\includegraphics[scale=.3]{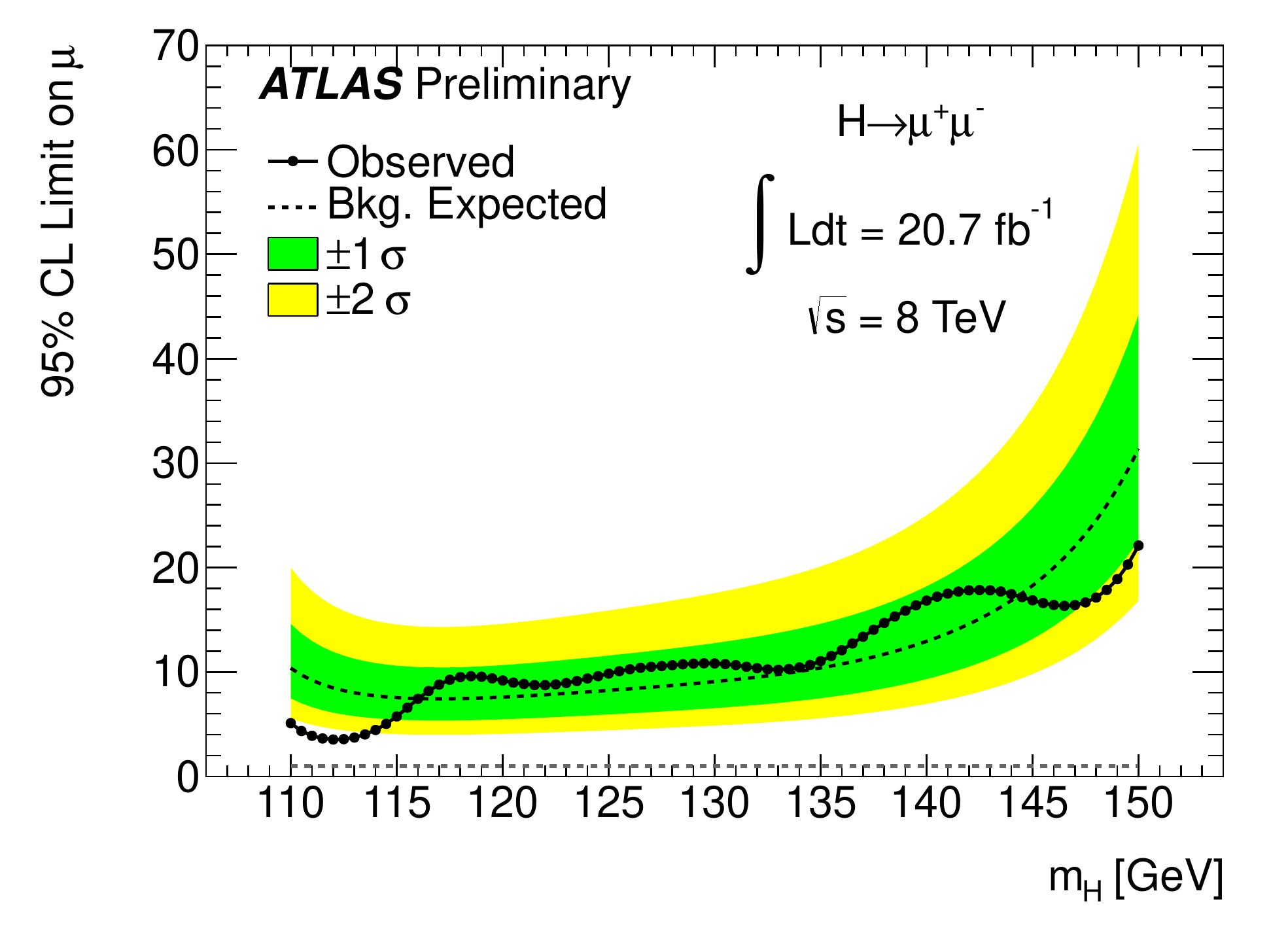}
\caption{\label{fig:hmumu_limit} Observed (solid) and expected (dashed) 95\% CL upper limits on the $H \rightarrow \mu \mu$ signal strength as a function of $m_{H}$ over the mass range $110 \, \mathrm{GeV} <m_{H}<150 \, \mathrm{GeV}$. The green and yellow regions indicate the $\pm 1 \, \sigma$ and $\pm \, 2 \sigma$ uncertainty bands on the expected limit, respectively \cite{ATLAS-CONF-2013-010}.}
\end{figure}

The $H \rightarrow \mu^{+} \mu^{-}$ decay is extremely rare, $BR = 28\times10^{-5} \, - \, 6 \times 10^{-5}$ over the $110 - 150 \, \mathrm{GeV}$ mass range, but it is the only channel were couplings to second generation fermions can be tested.
This channel benefits from a clear di-muon signature, which allows for an excellent mass resolution, but is subject to dominant irreducible Drell-Yan background contamination. Two analysis categories, depending on the muons detector rapidity region were considered. 
A binned likelihood fit to the data was performed using analytical shapes to model signal and background contributions.
No excess above the SM background expectation was observed. Limits were therefore set, which are shown in Figure \ref{fig:hmumu_limit}. The limit at $m_{H} = 125 \, \mathrm{GeV}$ on $\sigma_{SM}\times BR =  9.8$ (8.2 expected).

\subsection{\label{sec:mass_combination} Mass measurement combination}
The mass of the observed boson was measured from the high mass resolution two-photon and four-lepton channels. 
The mass systematic uncertainty is dominated in the two-photon channel by photon energy scale systematics and by the muon momentum scale in the four-lepton channel.
The best measured mass is $m_{H} = 125.5 \pm 0.2 \, \mathrm{(stat)} ^{+0.5}_{-0.6} \, \mathrm{(syst) \, GeV}$.
The mass difference between the two-photon and four-lepton channel is $\Delta m_{H} = 2.3^{+0.6}_{-0.7} \, \mathrm{(stat)} \pm 0.6 \, \mathrm{(syst) \, GeV}$, which is $2.4 \, \sigma$ away from the expected value of 0. The probability for such an occurrence is p = 1.5\%. If rectangular probability distribution functions were used for the energy and momentum scale systematic uncertainties p = 8\%.

\begin{figure}[h]
\includegraphics[scale=.3]{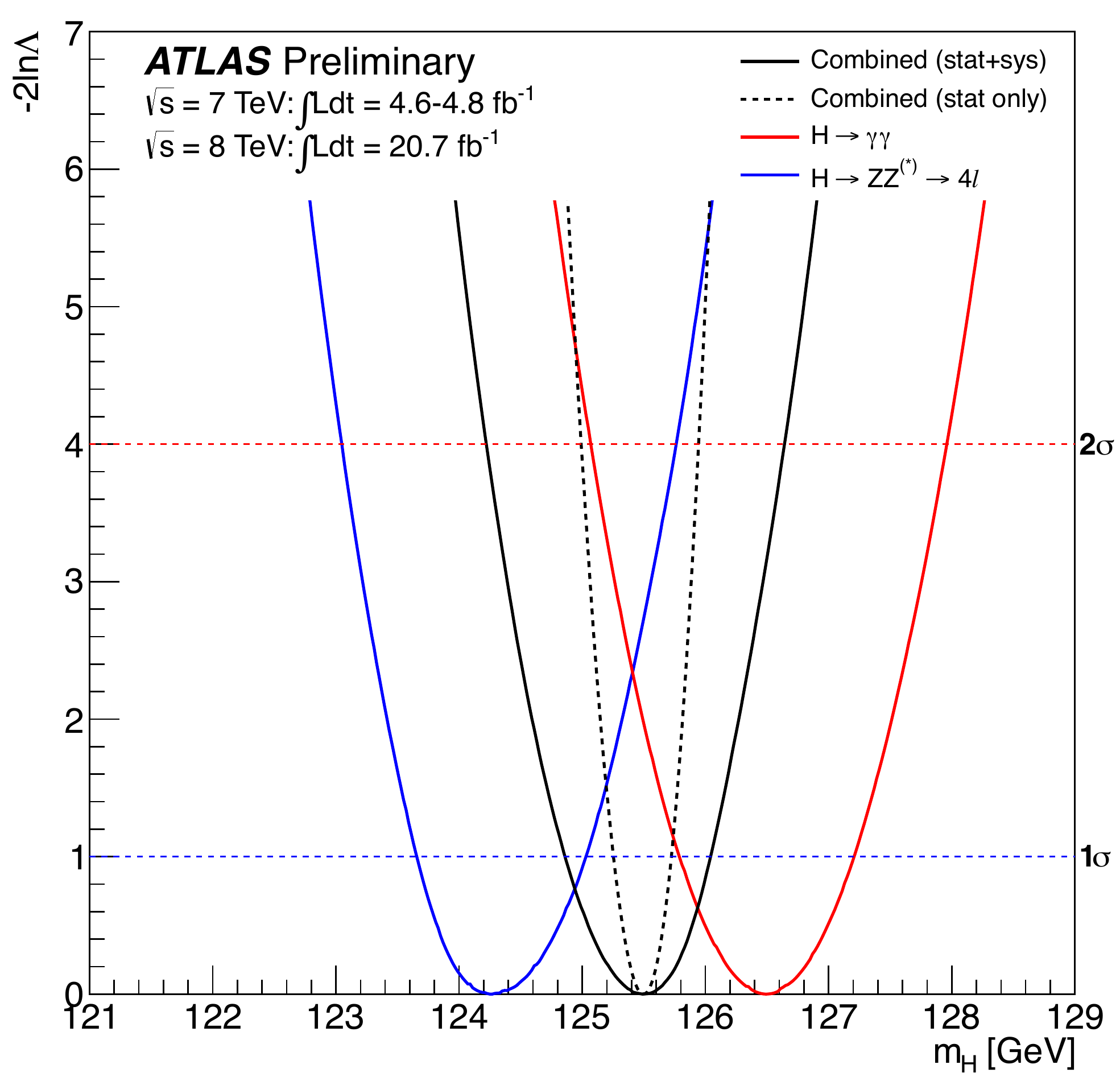}
\caption{\label{fig:combination_mass}The profile likelihood ratio $-2\ln{\Lambda(m_{H})}$ as a function of $m_{H}$ for the two-photon and four-lepton channels and their combination. The dashed line shows the statistical component of the mass measurement uncertainty \cite{ATLAS-CONF-2013-014}.}
\end{figure}

\subsection{\label{sec:signal_strenght_combination} Signal strength combination}

\begin{figure}[h]
\includegraphics[scale=.15]{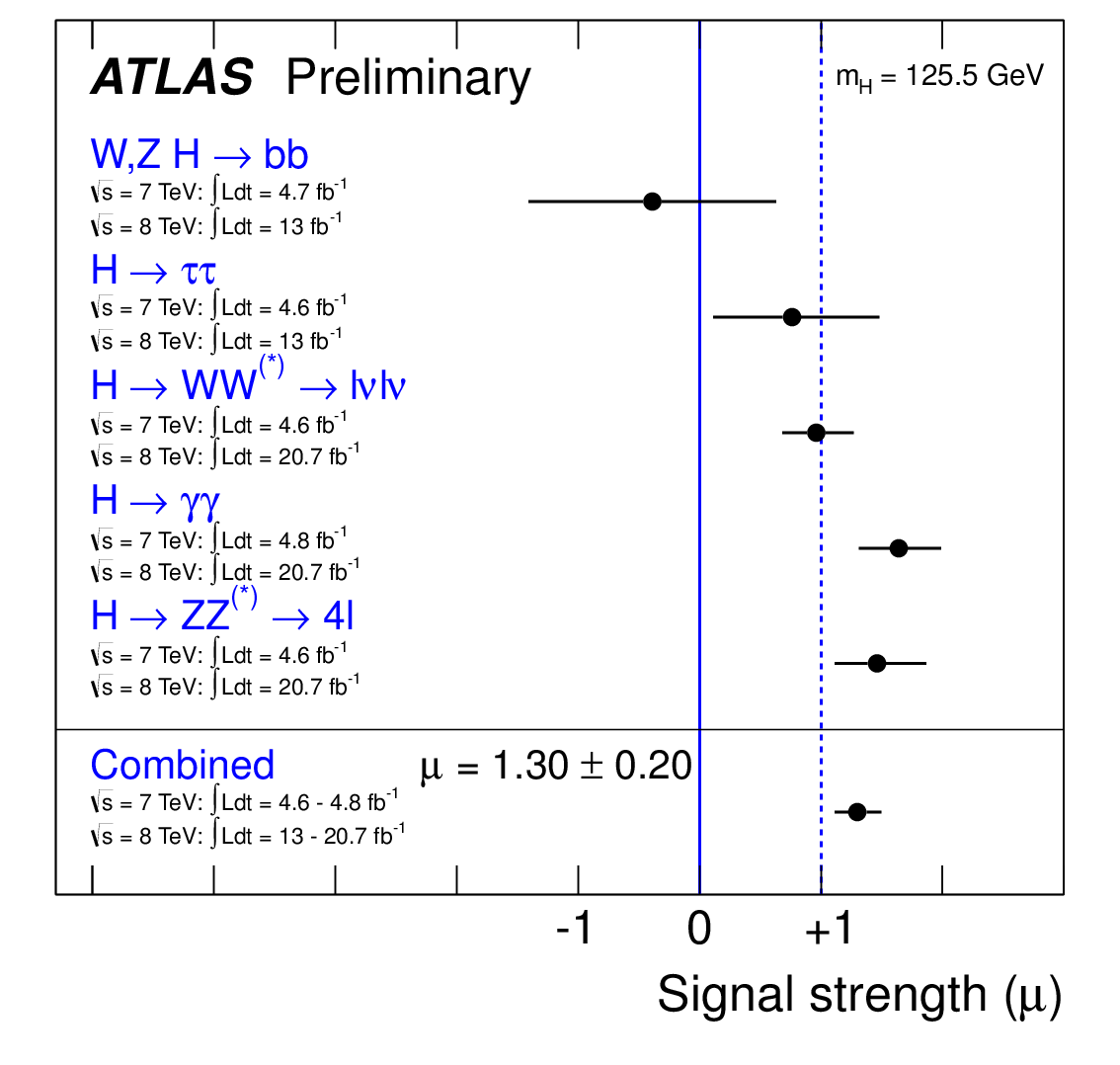}
\caption{\label{fig:couplings_new_signal_strenghts} Measurements of the signal strength parameter $\mu$ for $m_{H} =125.5 \, \mathrm{GeV}$ for the individual channels and for their combination \cite{ATLAS-CONF-2013-014}.}
\end{figure}

For the combination of the signal strength, besides the high mass resolution channels, the $WW$, $\tau \tau$ and $b\overline{b}$ channels were included.
The best measured signal strenght is $\mu = 1.30 \pm 0.13 \, \mathrm{(stat)} \pm 0.14 \, \mathrm{(sys)}$. Figure \ref{fig:couplings_new_signal_strenghts} shows the signal strengths for the individual channels and the combined one.  

The consistency of the combined measurement of $\mu$ with the SM expectation is of 9\%, and rises to 40\% if rectangular probability density functions were used for parton distribution functions and QCD scale uncertainties. The consistency of the individual $\mu$ measurements with the SM is 8\%, and 13\% with the $\mu = 1.43$ value. 

\subsection{\label{sec:properties_combination} Couplings combination}
\begin{figure}[h]
\includegraphics[scale=.15]{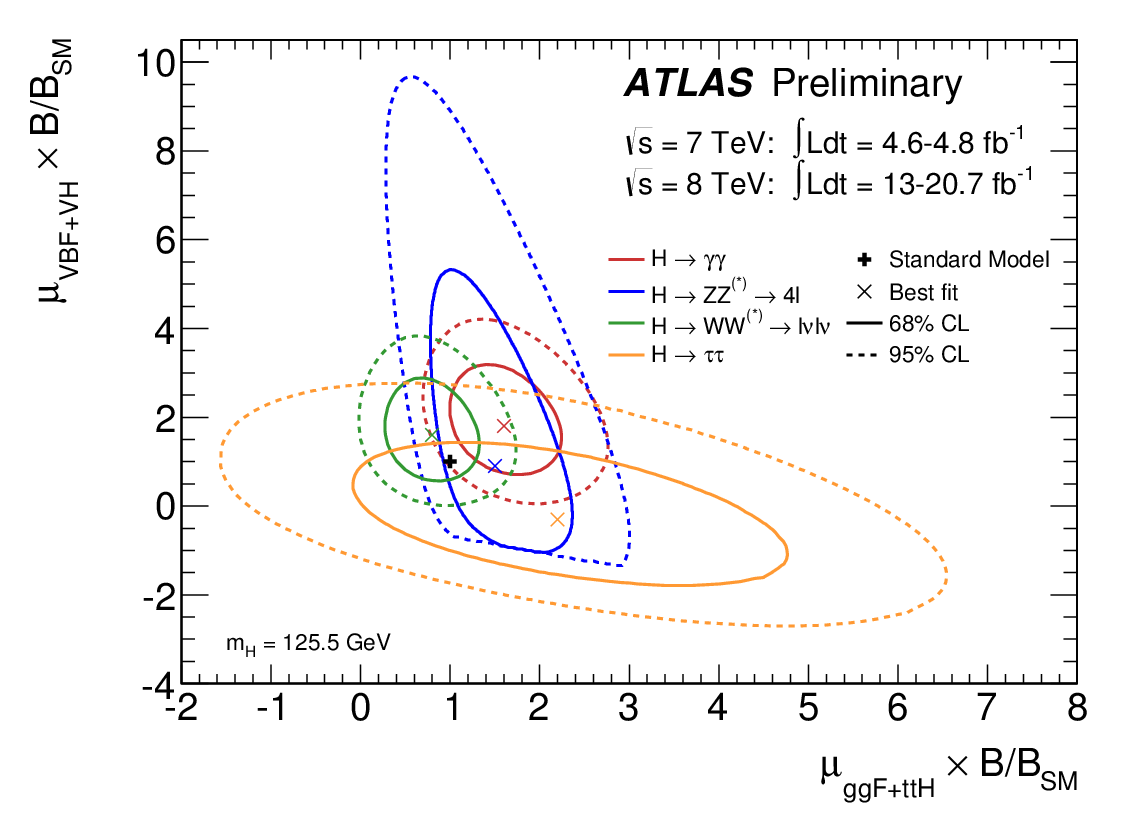}
\caption{\label{fig:couplings_new_production_modes} Likelihood contours for the two-photon, four-lepton, $WW$ and $\tau \tau$ channels in the $(\mu_{\mathrm{ggF+t\overline{t}H}}, \mu_{\mathrm{VBF+VH}})$ plane for a Higgs boson mass hypothesis of $m_{H} = 125.5 \, \mathrm{GeV}$. The best-fit to the data ($\times$) and 68\% (full) and 95\% (dashed) CL contours are also indicated, as well as the SM expectation ($+$) \cite{ATLAS-CONF-2013-034}.}
\end{figure}

More tests were performed on the coupling properties of the new boson. 
Two separate signal strengths for the $\mathrm{ggF+t\overline{t}H}$ and $\mathrm{VBF+VH}$ production modes were fitted. The results are shown in Figure \ref{fig:couplings_new_production_modes}. The signal strength ratio $\mu_{\mathrm{VBF+VH}}/\mu_{\mathrm{ggF+t\overline{t}H}} = 1.2^{+0.7}_{-0.5}$, providing a $3 \, \sigma$ evidence for production through vector-boson fusion.

\begin{figure}[h]
\includegraphics[scale=.15]{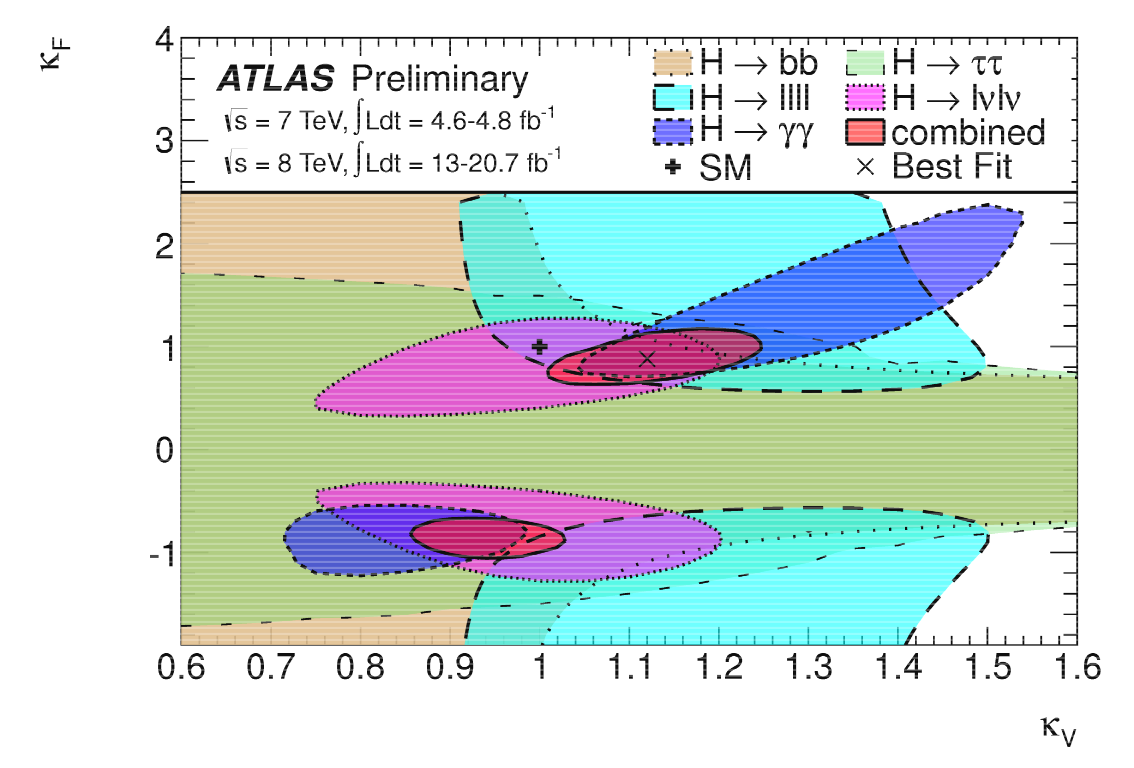}
\caption{\label{fig:couplings_new_fermion_boson}Fit for 2-parameter benchmark model probing different coupling strength scale factors for fermions and vector bosons, assuming only SM contributions to the total width \cite{ATLAS-CONF-2013-034}.}
\end{figure}

An analysis of vector boson and fermion couplings was performed under the assumption that couplings scale factors with respect to the SM are the same for vector bosons and fermions respectively. 
The results under the assumption that only SM particles are involved are shown in Figure \ref{fig:couplings_new_fermion_boson}. The compatibility with the SM is 8\%.

\begin{figure}[h!]
\includegraphics[scale=.15]{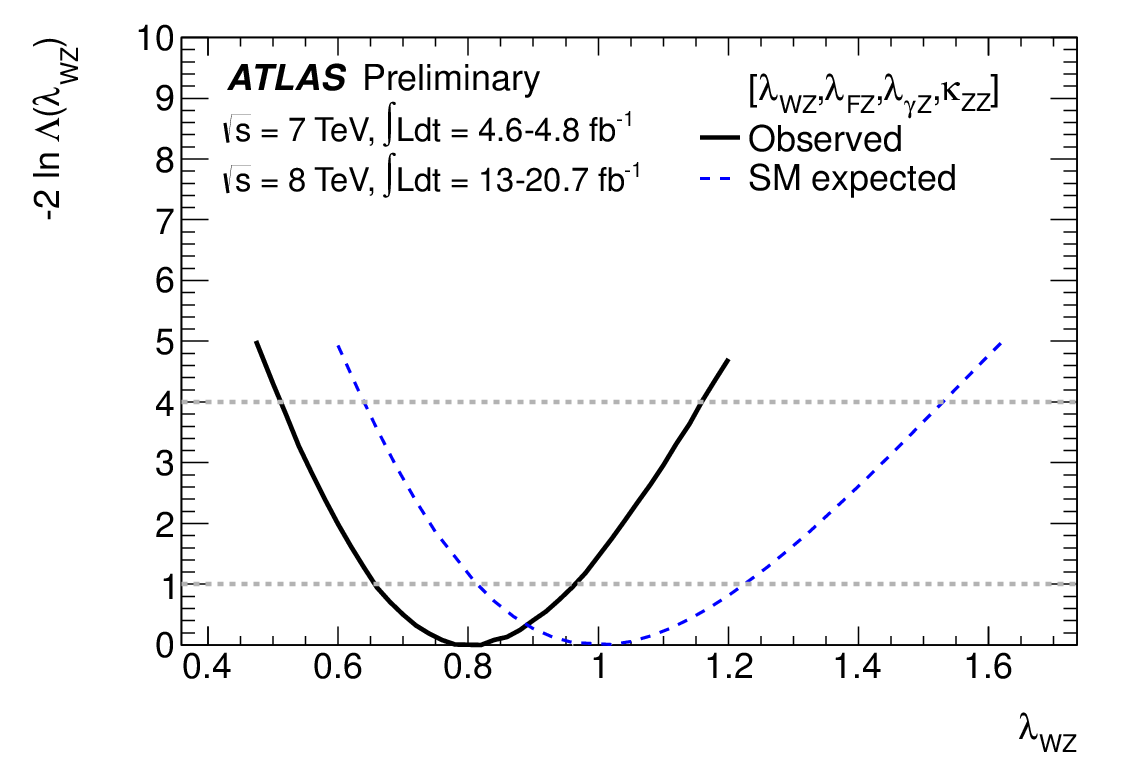}
\caption{\label{fig:couplings_new_custodial_noggassumpt}Fits for the benchmark model probing the custodial symmetry. The dashed curve shows the SM expectation \cite{ATLAS-CONF-2013-034}.}
\end{figure}

Custodial symmetry was analysed taking as only free parameter the ratio of scale factors for the couplings to the $W$ and $Z$ vector bosons. Results are shown in Figure \ref{fig:couplings_new_custodial_noggassumpt} and show a good agreement with the SM expectation at the 95\% CL. 

Finally possible contributions from non SM particles to the $gg \rightarrow H$ and $H \rightarrow \gamma \gamma$ loops were considered. 

\clearpage

\begin{figure}[h!]
\includegraphics[scale=.15]{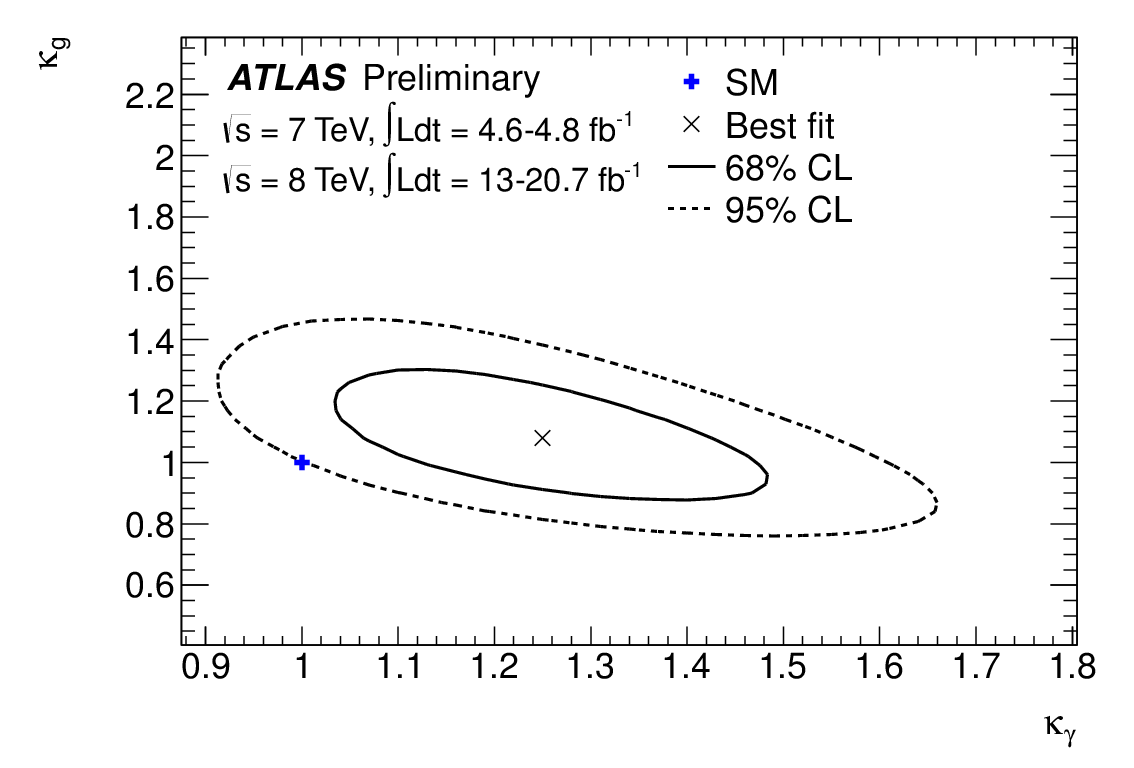}
\caption{\label{fig:couplings_new_new_particles_fixed_width} Fit for the benchmark model probing contributions from non-SM particles in the $H \rightarrow \gamma \gamma$ and $gg \rightarrow H$ loops. The dashed curve shows the SM expectation \cite{ATLAS-CONF-2013-034}.}
\end{figure}

\noindent
Results assuming no sizeable extra contributions to the total width are shown in Figure \ref{fig:couplings_new_new_particles_fixed_width}. There is no evidence of contributions from non SM effects.  

\section{\label{sec:bsm_searches} Beyond the Standard Model searches}
In the following two different types of BSM searches are briefly summarised: a generic search for an invisibly decaying Higgs and two Minimal Supersymmetric Standard Model (MSSM) searches. 
In the MSSM the minimal realisation of the Higgs sector requires the introduction of two Higgs doublets, which correspond to five physical states, three neutrals and two charged. The Higgs sector is then typically described by the mass of the neutral CP-odd $A$ boson and $\tan{\beta}$, which is the ratio of the two doublets vacuum expectation values.

\subsection{\label{sec:invisible} Invisible channel}
\begin{figure}[h]
   \subfigure[]{ 
        \begin{overpic}[width=0.3\textwidth]{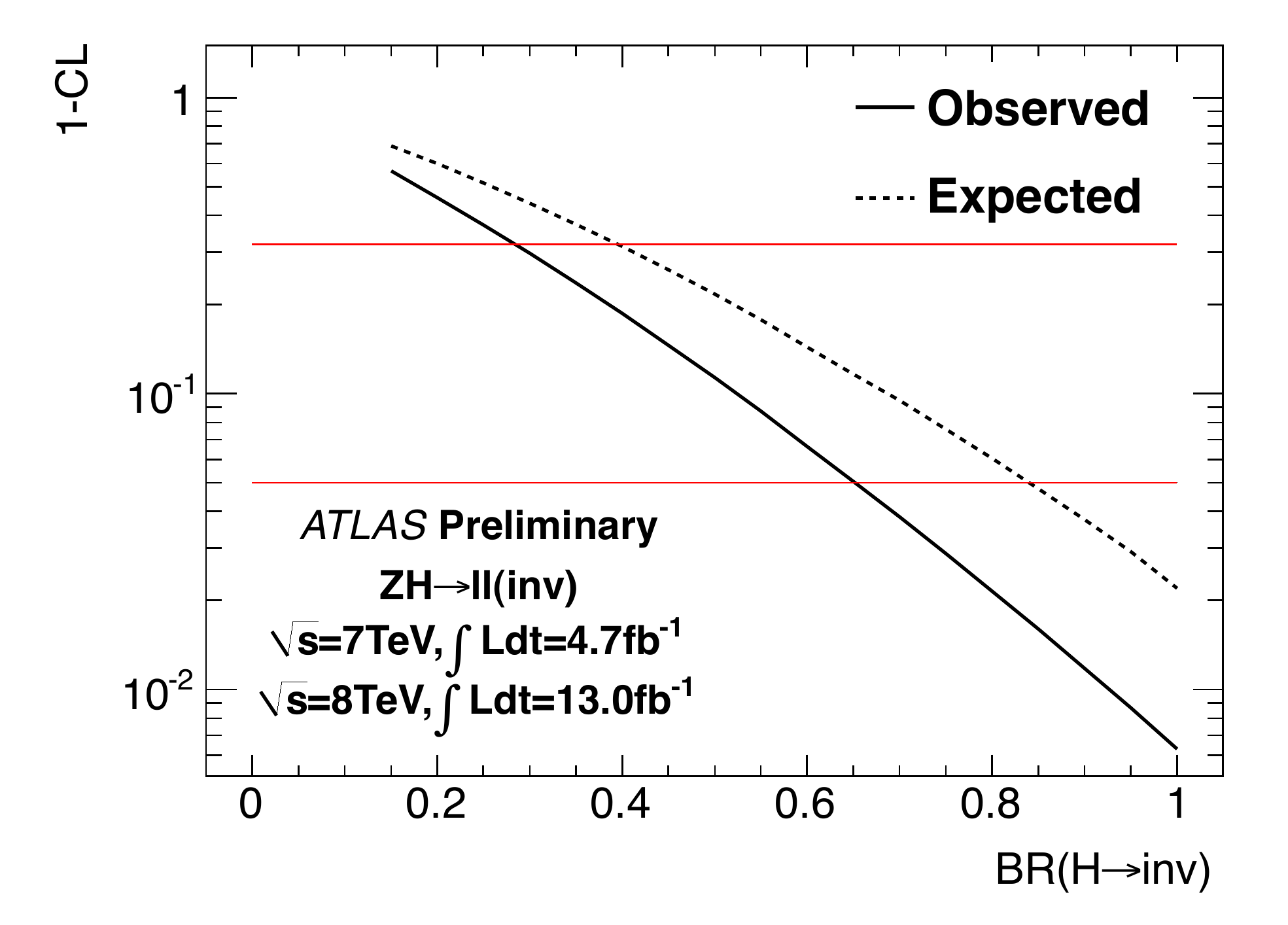}
        \label{fig:invisible_br_exclusion}
        \end{overpic}
    }
    
      \subfigure[]{ 
        \begin{overpic}[width=0.3\textwidth]{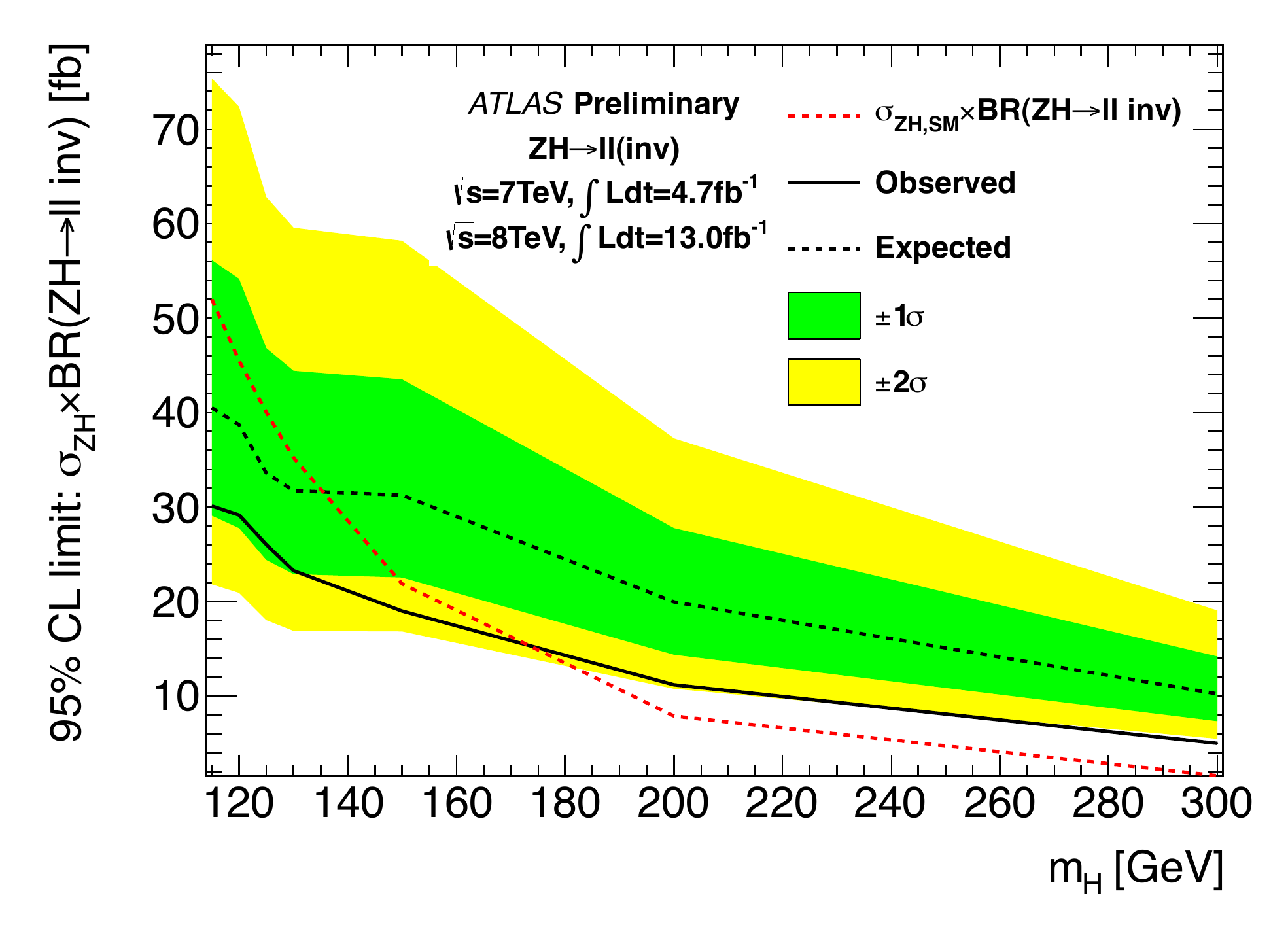}
        \label{fig:invisible_limit}
        \end{overpic}
    }  
    
     \caption{Interpretations of the results of the $H \rightarrow \, \mathrm{invisible}$ search. (a) Confidence level scanned against $BR(H \rightarrow \, \mathrm{invisible})$ for the SM Higgs boson with $m_{H} = 125 \, \mathrm{GeV}$. The dashed line shows the expected values, whereas the solid line indicates the observed values. The red solid lines indicate the 68\% and 95\% CL. (b) 95\% confidence level limits on the cross-section times branching fraction of a Higgs-like boson decaying to invisible particles. Dashed lines show the background only expected limits and solid lines show the observed limit \cite{ATLAS-CONF-2013-011}.}
     
\end{figure}

In the SM the Higgs branching ratio to invisible particles is not measurable. This branching ratio could however have sizeable contributions from BSM processes, for example from dark matter particles. 
The signature is given by a $Z$ boson and large missing transverse momentum. In the analysis $Z$ decays to electrons and muons were considered.
The main backgrounds come from di-boson production. No excess above the SM background expectation was observed. Results are shown in Figures \ref{fig:invisible_br_exclusion} and \ref{fig:invisible_limit} according to two different interpretations. 
A limit is set for the invisible branching ratio of a SM Higgs boson \ref{fig:invisible_br_exclusion} and one on the product of $ZH$ production cross-section and invisible decay branching ratio for further Higgs-like states  \ref{fig:invisible_limit}.

\subsection{\label{sec:invisible} MSSM neutral Higgs}
\begin{figure}[t]
\includegraphics[scale=.3]{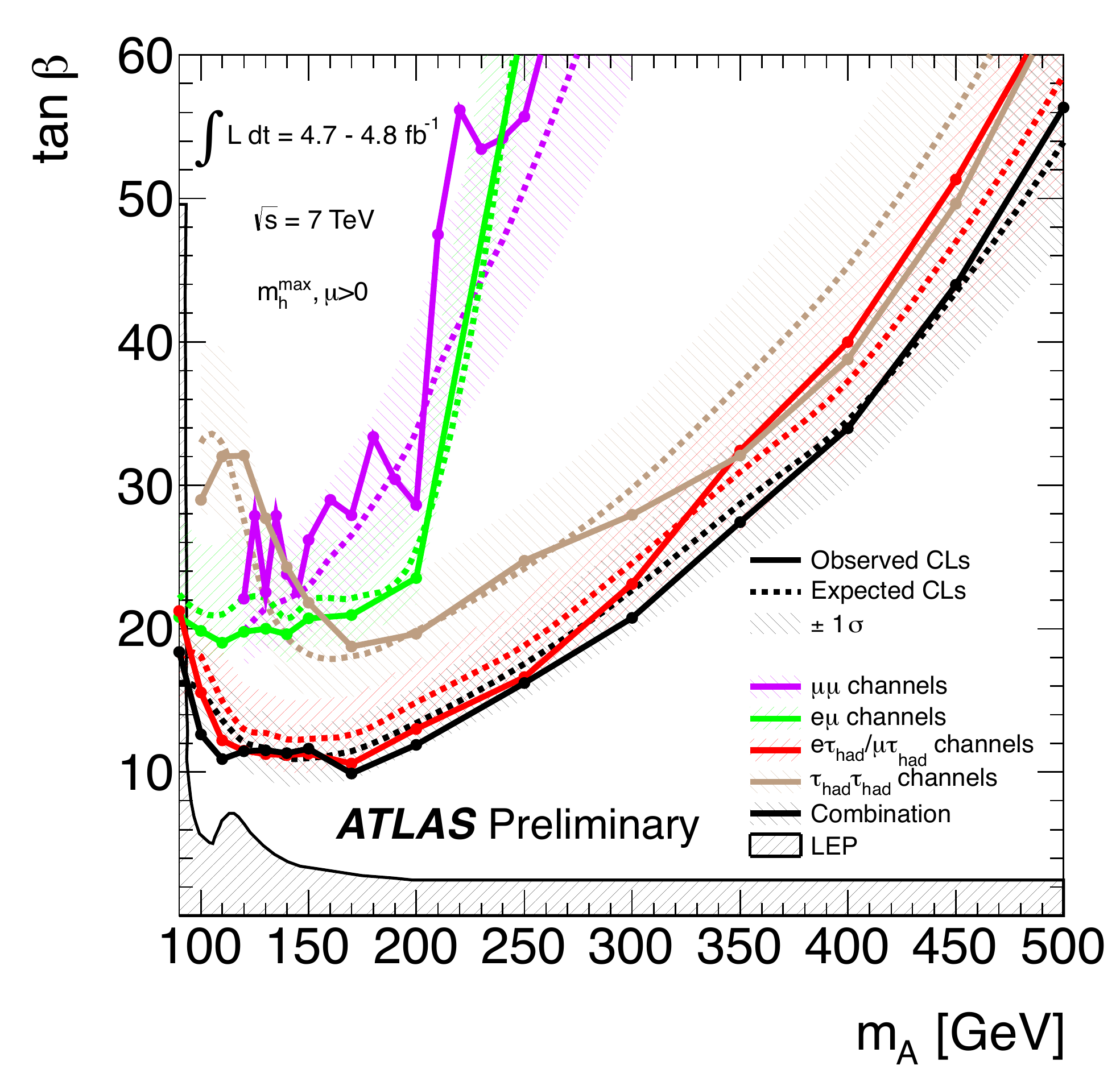}
\caption{\label{fig:mssm_neutral_limit}Expected (dashed line) and observed (solid line) 95\% confidence level CL limits on $\tan\beta$ as a function of $m_{A}$ for individual final states and for the combination \cite{ATLAS-CONF-2012-094}.}
\end{figure}

In the MSSM the neutral Higgs states branching ratios have a structure similar to that of the SM, but suppressed or enhanced according to $\tan \beta$. 
At high $\tan \beta$ decays of A and H to tau leptons and muons are highly favoured.
In the search performed the following final states were considered, where $\tau_{\mathrm{had}}$ denotes hadronic tau decays: $e\mu$,  $e \tau_{\mathrm{had}}$, $\mu \tau_{\mathrm{had}}$, $\tau_{\mathrm{had}}\tau_{\mathrm{had}}$.
The two main MSSM production modes, ggF and $b$-quark associated production were exploited in the analysis.
No excess above the SM expectation was observed. Limits were therefore set in the $(m_{A}, \tan{\beta})$ plane, that are shown in Figure \ref{fig:mssm_neutral_limit}.

\subsection{\label{sec:invisible} MSSM charged Higgs}
\begin{figure}[t]
\includegraphics[scale=.3]{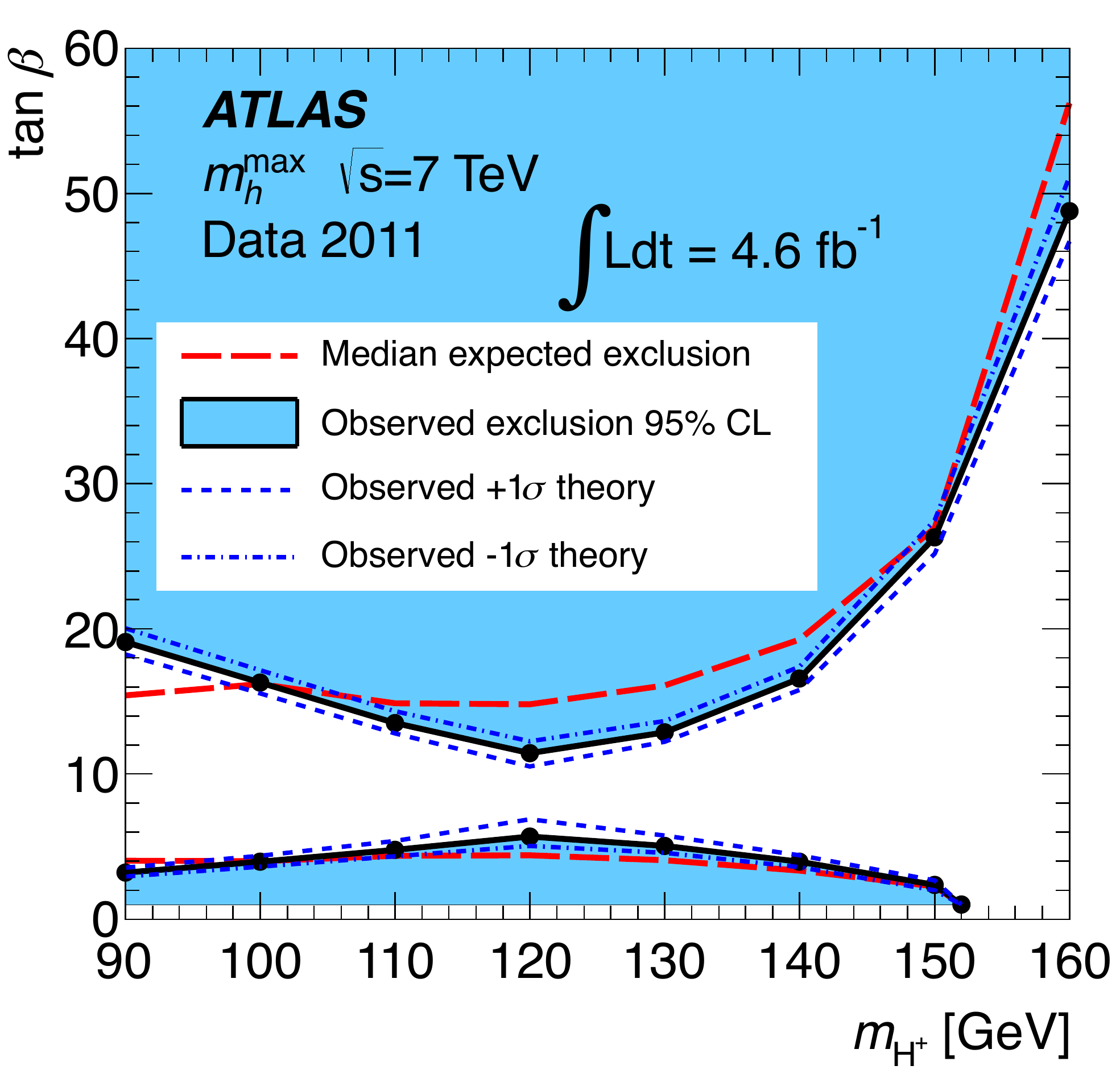}
\caption{\label{fig:MSSM_charged_limit}Combined 95\% CL exclusion limits on $\tan{\beta}$ as a function of $m_{H^{+}}$ \cite{Aad:2012tj}.}
\end{figure}

Since at least two Higgs doublets are needed in supersymmetric scenarios, the search for charged Higgs states is of great importance. The main production mode for the charged Higgs in the MSSM is through top decays. The main decay mode for the charged Higgs is $H^{\pm} \rightarrow \tau \nu$. 
Three final states were considered in the search, one where the $W$ from $t \rightarrow Wb$ decays hadronically and the tau to leptons, one with the $W$ decaying leptonically and the tau hadronically and finally one where both the $W$ and the tau decay to hadrons.
The background is mainly due to $t\overline{t}$ pairs, single-top quark production, multi-jet and di-boson events.
A mixed MC and data-driven background estimation strategy was used.
Since no excess above the SM expectation was observed, limits in the $(m_{H^{+}}, \tan{\beta})$ plane were set and are shown in Figure~\ref{fig:MSSM_charged_limit}.

\section{\label{sec:conclusion} Conclusion}
Since the announcement of the discovery of a Higgs-like boson measurements and searches for other decay channels or other states have proceeded intensively. 
The observation of the new boson is well established in the two-photon, four-lepton and $WW$ channels, with $7.4 \, \sigma, 6.6 \, \sigma, 3.8 \, \sigma$ significances respectively.
The combined mass measurement provides a best measured mass of the new boson $m_H = 125.5 \pm 0.2 \, \mathrm{(stat)}^{+0.5}_{-0.6} \, \mathrm{(syst) \, GeV}$.
Fermionic decays were searched for. Limits were set  at $1.9 \times \sigma_{SM}$and $1.8 \times \sigma_{SM}$ at $m_{H} = 125 \, \mathrm{GeV}$ for the $\tau \tau$ and $b\overline{b}$ channels respectively. The full LHC datasets have not yet been employed for these channels.
The combined signal strength is $\mu = 1.30 \pm 0.13 \, \mathrm{(stat)} \pm 0.14 \, \mathrm{(sys)}$.
Results for couplings and spin-parity properties are compatible with the SM expectations.
No evidence for invisible Higgs decays or BSM Higgs states was observed.
The characteristics of the observed boson are up to now compatible with those of a SM Higgs boson.

\begin{acknowledgments}
I thank first of all the LHC and ATLAS teams, for the operation of the LHC and of the ATLAS detector, as well as the institutions supporting ATLAS. I thank in particular the INFN for the funding support.
I would like to thank the ATLAS speakers committee and the ATLAS Higgs group for the opportunity of giving this talk on behalf of the collaboration. I thank Marumi Kado and my supervisors Donatella Cavalli and Attilio Andreazza for the help and advice in preparing for this conference.
I finally thank the organizers of LISHEP for the opportunity to come to Rio de Janeiro for this interesting conference.
\end{acknowledgments}

\nocite{*}

\bibliography{lishep_proc_consonni}

\begin{thebibliography}{22}%
\makeatletter
\providecommand \@ifxundefined [1]{%
 \@ifx{#1\undefined}
}%
\providecommand \@ifnum [1]{%
 \ifnum #1\expandafter \@firstoftwo
 \else \expandafter \@secondoftwo
 \fi
}%
\providecommand \@ifx [1]{%
 \ifx #1\expandafter \@firstoftwo
 \else \expandafter \@secondoftwo
 \fi
}%
\providecommand \natexlab [1]{#1}%
\providecommand \enquote  [1]{``#1''}%
\providecommand \bibnamefont  [1]{#1}%
\providecommand \bibfnamefont [1]{#1}%
\providecommand \citenamefont [1]{#1}%
\providecommand \href@noop [0]{\@secondoftwo}%
\providecommand \href [0]{\begingroup \@sanitize@url \@href}%
\providecommand \@href[1]{\@@startlink{#1}\@@href}%
\providecommand \@@href[1]{\endgroup#1\@@endlink}%
\providecommand \@sanitize@url [0]{\catcode `\\12\catcode `\$12\catcode
  `\&12\catcode `\#12\catcode `\^12\catcode `\_12\catcode `\%12\relax}%
\providecommand \@@startlink[1]{}%
\providecommand \@@endlink[0]{}%
\providecommand \url  [0]{\begingroup\@sanitize@url \@url }%
\providecommand \@url [1]{\endgroup\@href {#1}{\urlprefix }}%
\providecommand \urlprefix  [0]{URL }%
\providecommand \Eprint [0]{\href }%
\providecommand \doibase [0]{http://dx.doi.org/}%
\providecommand \selectlanguage [0]{\@gobble}%
\providecommand \bibinfo  [0]{\@secondoftwo}%
\providecommand \bibfield  [0]{\@secondoftwo}%
\providecommand \translation [1]{[#1]}%
\providecommand \BibitemOpen [0]{}%
\providecommand \bibitemStop [0]{}%
\providecommand \bibitemNoStop [0]{.\EOS\space}%
\providecommand \EOS [0]{\spacefactor3000\relax}%
\providecommand \BibitemShut  [1]{\csname bibitem#1\endcsname}%
\let\auto@bib@innerbib\@empty
\bibitem [{\citenamefont {Englert}\ and\ \citenamefont
  {Brout}(1964)}]{PhysRevLett.13.321}%
  \BibitemOpen
  \bibfield  {author} {\bibinfo {author} {\bibfnamefont {F.}~\bibnamefont
  {Englert}}\ and\ \bibinfo {author} {\bibfnamefont {R.}~\bibnamefont
  {Brout}},\ }\href {\doibase 10.1103/PhysRevLett.13.321} {\bibfield  {journal}
  {\bibinfo  {journal} {Phys. Rev. Lett.}\ }\textbf {\bibinfo {volume} {13}},\
  \bibinfo {pages} {321} (\bibinfo {year} {1964})}\BibitemShut {NoStop}%
\bibitem [{\citenamefont {Higgs}(1964)}]{PhysRevLett.13.508}%
  \BibitemOpen
  \bibfield  {author} {\bibinfo {author} {\bibfnamefont {P.~W.}\ \bibnamefont
  {Higgs}},\ }\href {\doibase 10.1103/PhysRevLett.13.508} {\bibfield  {journal}
  {\bibinfo  {journal} {Phys. Rev. Lett.}\ }\textbf {\bibinfo {volume} {13}},\
  \bibinfo {pages} {508} (\bibinfo {year} {1964})}\BibitemShut {NoStop}%
\bibitem [{\citenamefont {Guralnik}\ \emph {et~al.}(1964)\citenamefont
  {Guralnik}, \citenamefont {Hagen},\ and\ \citenamefont
  {Kibble}}]{PhysRevLett.13.585}%
  \BibitemOpen
  \bibfield  {author} {\bibinfo {author} {\bibfnamefont {G.~S.}\ \bibnamefont
  {Guralnik}}, \bibinfo {author} {\bibfnamefont {C.~R.}\ \bibnamefont {Hagen}},
  \ and\ \bibinfo {author} {\bibfnamefont {T.~W.~B.}\ \bibnamefont {Kibble}},\
  }\href {\doibase 10.1103/PhysRevLett.13.585} {\bibfield  {journal} {\bibinfo
  {journal} {Phys. Rev. Lett.}\ }\textbf {\bibinfo {volume} {13}},\ \bibinfo
  {pages} {585} (\bibinfo {year} {1964})}\BibitemShut {NoStop}%
\bibitem [{\citenamefont {{The ATLAS
  Collaboration}}(2012{\natexlab{a}})}]{Aad20121}%
  \BibitemOpen
  \bibfield  {author} {\bibinfo {author} {\bibnamefont {{The ATLAS
  Collaboration}}},\ }\href {\doibase 10.1016/j.physletb.2012.08.020}
  {\bibfield  {journal} {\bibinfo  {journal} {Physics Letters B}\ }\textbf
  {\bibinfo {volume} {716}},\ \bibinfo {pages} {1 } (\bibinfo {year}
  {2012}{\natexlab{a}})}\BibitemShut {NoStop}%
\bibitem [{\citenamefont {{The CMS Collaboration}}(2012)}]{Chatrchyan:1471016}%
  \BibitemOpen
  \bibfield  {author} {\bibinfo {author} {\bibnamefont {{The CMS
  Collaboration}}},\ }\href@noop {} {\bibfield  {journal} {\bibinfo  {journal}
  {Phys. Lett. B}\ }\textbf {\bibinfo {volume} {716}},\ \bibinfo {pages} {30}
  (\bibinfo {year} {2012})}\BibitemShut {NoStop}%
\bibitem [{\citenamefont {Dittmaier}\ \emph {et~al.}(2011)\citenamefont
  {Dittmaier} \emph {et~al.}}]{Dittmaier:2011ti}%
  \BibitemOpen
  \bibfield  {author} {\bibinfo {author} {\bibfnamefont {S.}~\bibnamefont
  {Dittmaier}} \emph {et~al.} (\bibinfo {collaboration} {LHC Higgs Cross
  Section Working Group}),\ }\href@noop {} {\  (\bibinfo {year} {2011})},\
  \Eprint {http://arxiv.org/abs/1101.0593} {arXiv:1101.0593 [hep-ph]}
  \BibitemShut {NoStop}%
\bibitem [{\citenamefont {{The ATLAS Collaboration}}(2008)}]{Aad:2008zzm}%
  \BibitemOpen
  \bibfield  {author} {\bibinfo {author} {\bibnamefont {{The ATLAS
  Collaboration}}},\ }\href@noop {} {\bibfield  {journal} {\bibinfo  {journal}
  {JINST}\ }\textbf {\bibinfo {volume} {3}},\ \bibinfo {pages} {S08003}
  (\bibinfo {year} {2008})}\BibitemShut {NoStop}%
\bibitem [{\citenamefont {{ATLAS
  Collaboration}}(2013{\natexlab{a}})}]{ATLAS-CONF-2013-014}%
  \BibitemOpen
  \bibfield  {author} {\bibinfo {author} {\bibnamefont {{ATLAS
  Collaboration}}},\ }\ \bibinfo {number} {ATLAS-CONF-2013-014},\
  {https://cds.cern.ch/record/1523727/}\BibitemShut {NoStop}%
\bibitem [{\citenamefont {{ATLAS
  Collaboration}}(2013{\natexlab{b}})}]{ATLAS-CONF-2013-012}%
  \BibitemOpen
  \bibfield  {author} {\bibinfo {author} {\bibnamefont {{ATLAS
  Collaboration}}},\ }\ \bibinfo {number} {ATLAS-CONF-2013-012},\
  {https://cds.cern.ch/record/1523698}\BibitemShut {NoStop}%
\bibitem [{\citenamefont {{ATLAS
  Collaboration}}(2013{\natexlab{c}})}]{ATLAS-CONF-2013-013}%
  \BibitemOpen
  \bibfield  {author} {\bibinfo {author} {\bibnamefont {{ATLAS
  Collaboration}}},\ }\ \bibinfo {number} {ATLAS-CONF-2013-013},\
  {https://cds.cern.ch/record/1523699/}\BibitemShut {NoStop}%
\bibitem [{\citenamefont {{ATLAS
  Collaboration}}(2013{\natexlab{d}})}]{ATLAS-CONF-2013-030}%
  \BibitemOpen
  \bibfield  {author} {\bibinfo {author} {\bibnamefont {{ATLAS
  Collaboration}}},\ }\ \bibinfo {number} {ATLAS-CONF-2013-030},\
  {https://cds.cern.ch/record/1527126}\BibitemShut {NoStop}%
\bibitem [{\citenamefont {{ATLAS
  Collaboration}}(2012{\natexlab{a}})}]{ATLAS-CONF-2012-161}%
  \BibitemOpen
  \bibfield  {author} {\bibinfo {author} {\bibnamefont {{ATLAS
  Collaboration}}},\ }\ \bibinfo {number} {ATLAS-CONF-2012-161},\
  {https://cds.cern.ch/record/1493625}\BibitemShut {NoStop}%
\bibitem [{\citenamefont {{ATLAS
  Collaboration}}(2012{\natexlab{b}})}]{ATLAS-CONF-2012-135}%
  \BibitemOpen
  \bibfield  {author} {\bibinfo {author} {\bibnamefont {{ATLAS
  Collaboration}}},\ }\ \bibinfo {number} {ATLAS-CONF-2012-135},\
  {https://cds.cern.ch/record/1478423}\BibitemShut {NoStop}%
\bibitem [{\citenamefont {{ATLAS
  Collaboration}}(2012{\natexlab{c}})}]{ATLAS-CONF-2012-160}%
  \BibitemOpen
  \bibfield  {author} {\bibinfo {author} {\bibnamefont {{ATLAS
  Collaboration}}},\ }\ \bibinfo {number} {ATLAS-CONF-2012-160},\
  {https://cds.cern.ch/record/1493624}\BibitemShut {NoStop}%
\bibitem [{\citenamefont {{ATLAS
  Collaboration}}(2013{\natexlab{e}})}]{ATLAS-CONF-2013-034}%
  \BibitemOpen
  \bibfield  {author} {\bibinfo {author} {\bibnamefont {{ATLAS
  Collaboration}}},\ }\ \bibinfo {number} {ATLAS-CONF-2013-034},\
  {https://cds.cern.ch/record/1528170}\BibitemShut {NoStop}%
\bibitem [{\citenamefont {{ATLAS
  Collaboration}}(2013{\natexlab{f}})}]{ATLAS-CONF-2013-029}%
  \BibitemOpen
  \bibfield  {author} {\bibinfo {author} {\bibnamefont {{ATLAS
  Collaboration}}},\ }\ \bibinfo {number} {ATLAS-CONF-2013-029},\
  {https://cds.cern.ch/record/1527124}\BibitemShut {NoStop}%
\bibitem [{\citenamefont {{ATLAS
  Collaboration}}(2013{\natexlab{g}})}]{ATLAS-CONF-2013-031}%
  \BibitemOpen
  \bibfield  {author} {\bibinfo {author} {\bibnamefont {{ATLAS
  Collaboration}}},\ }\ \bibinfo {number} {ATLAS-CONF-2013-031},\
  {https://cds.cern.ch/record/1527127}\BibitemShut {NoStop}%
\bibitem [{\citenamefont {{ATLAS
  Collaboration}}(2013{\natexlab{h}})}]{ATLAS-CONF-2013-009}%
  \BibitemOpen
  \bibfield  {author} {\bibinfo {author} {\bibnamefont {{ATLAS
  Collaboration}}},\ }\ \bibinfo {number} {ATLAS-CONF-2013-009},\
  {https://cds.cern.ch/record/1523683}\BibitemShut {NoStop}%
\bibitem [{\citenamefont {{ATLAS
  Collaboration}}(2013{\natexlab{i}})}]{ATLAS-CONF-2013-010}%
  \BibitemOpen
  \bibfield  {author} {\bibinfo {author} {\bibnamefont {{ATLAS
  Collaboration}}},\ }\ \bibinfo {number} {ATLAS-CONF-2013-010},\
  {https://cds.cern.ch/record/1523695/}\BibitemShut {NoStop}%
\bibitem [{\citenamefont {{ATLAS
  Collaboration}}(2013{\natexlab{j}})}]{ATLAS-CONF-2013-011}%
  \BibitemOpen
  \bibfield  {author} {\bibinfo {author} {\bibnamefont {{ATLAS
  Collaboration}}},\ }\ \bibinfo {number} {ATLAS-CONF-2013-011},\
  {https://cds.cern.ch/record/1523696/}\BibitemShut {NoStop}%
\bibitem [{\citenamefont {{ATLAS
  Collaboration}}(2012{\natexlab{d}})}]{ATLAS-CONF-2012-094}%
  \BibitemOpen
  \bibfield  {author} {\bibinfo {author} {\bibnamefont {{ATLAS
  Collaboration}}},\ }\ \bibinfo {number} {ATLAS-CONF-2012-094},\
  {https://cds.cern.ch/record/1460440}\BibitemShut {NoStop}%
\bibitem [{\citenamefont {{The ATLAS
  Collaboration}}(2012{\natexlab{b}})}]{Aad:2012tj}%
  \BibitemOpen
  \bibfield  {author} {\bibinfo {author} {\bibnamefont {{The ATLAS
  Collaboration}}},\ }\href {\doibase 10.1007/JHEP06(2012)039} {\bibfield
  {journal} {\bibinfo  {journal} {JHEP}\ }\textbf {\bibinfo {volume} {1206}},\
  \bibinfo {pages} {039} (\bibinfo {year} {2012}{\natexlab{b}})},\ \Eprint
  {http://arxiv.org/abs/1204.2760} {arXiv:1204.2760 [hep-ex]} \BibitemShut
  {NoStop}%
\end{thebibliography}%

\end{document}